\documentclass[useAMS,usenatbib]{mn2e}
\usepackage{graphicx}
\def \th {\thinspace}
\def \erg {erg s$^{-1}$}
\def \arcmin {\hbox{$^\prime$}}

\def \degmark {^\circ}

\title[Unified model for LMXB]
{The nature of the Island and Banana States in Atoll\\ sources and a Unified Model for Low Mass X-ray Binaries}
\author[Church et al.]
{M. J. Church$^{1}$,  A. Gibiec$^{2}$ and M. Ba\l uci\'nska-Church$^{1}$\\
$^{1}$ School of Physics and Astronomy, University of Birmingham, Birmingham, B15 2TT, UK\\
$^{2}$ Astronomical Observatory, Jagiellonian University, ul. Orla 171, 30-244 Cracow, Poland.\\}
\begin{document}
\date{Accepted 2013 December 5.  Received 2013 December 5; in original form 2013 July 12}

\pagerange{\pageref{firstpage}--\pageref{lastpage}} \pubyear{2013}

\maketitle

\label{firstpage}

\begin{abstract}
We propose an explanation of the Island and Banana states and the relation between Atoll and Z-track 
sources, constituting a unified model for Low Mass X-ray Binaries (LMXB). We find a dramatic transition 
at a luminosity of 1 -- 2$\times 10^{37}$ erg s$^{-1}$ above which the high energy cut-off $E_{\rm CO}$ 
of the Comptonized emission in all sources is low at a few keV. There is thermal equilibrium between 
the neutron star at $\sim$2 keV and the Comptonizing accretion disk corona (ADC) causing the low 
$E_{\rm CO}$ in the Banana state of Atolls and all states of the Z-track sources. Below this luminosity, 
$E_{\rm CO}$ increases towards 100 keV causing the hardness of the island state. Thermal equilibrium is 
lost, the ADC becoming much hotter than the neutron star via an additional coronal heating mechanism.
This suggests a unified model of LMXB: the Banana state is a basic state with the mass accretion rate 
$\dot M$ increasing, corresponding to the Normal Branch of Z-track sources. The Island state has high ADC 
temperature, this state not existing in the Z-sources with luminosities much greater than the critical value. 
The Z-track sources have an additional flaring branch consistent with unstable nuclear burning on the 
neutron star at high $\dot M$. This burning r\'egime does not exist at low $\dot M$ so this branch is not 
seen in Atolls (except GX Atolls). The Horizontal Branch in Z-track sources has a strong increase in 
radiation pressure disrupting the inner disk and launching relativistic jets.
\end{abstract}


\section{Introduction}

The physical understanding of low mass X-ray binaries (LMXB) remains an important astrophysical problem.
We previously proposed an explanation of the brightest LMXB: the Z-track sources which convincingly
and consistently explained these sources and suggested a jet-launching mechanism (Church et al. 2006). 
In the present work we apply the same approach to the lower luminosity division of LMXB - the Atoll 
sources and show that this reveals the nature of these sources, and thus the nature of LMXB in general.

Low mass X-ray binaries span a wide range of luminosities from $\sim1\times 10^{36}$ to several 
$10^{38}$ \erg. Below $\sim10^{38}$ \erg, they display atoll source behaviour in which 
two main tracks are seen in hardness-intensity diagrams: the banana and island states (Fig. 1 left). In the 
banana state, hardness is relatively constant over a wide range of luminosity, while in the island 
state at the lowest luminosities the spectrum becomes harder (van der Klis 1995; Hasinger \& van der Klis 
1989). The banana state has been further divided into lower banana and upper banana at lower and
higher luminosities. Very faint sources exhibit an extreme island state (van Straaten et al. 2003) which
appears in hardness-intensity simply as the extreme end of the island state; however, in colour-colour 
it is seen as a separate track.

The most luminous sources with luminosities at or above the Eddington limit are the very different Z-track 
sources having three states evident from the Z-shapes they display in hardness-intensity: the Horizontal (HB), 
Normal (NB) and Flaring Branches (FB)  (Schulz et al. 1989; Hasinger \& van der Klis 1989). These indicate 
strong spectral changes reflecting marked physical changes within the sources (Fig. 1 right). 

Between the atoll and Z-track sources in luminosity lie the GX-atoll or bright atoll sources such as
GX\th 3+1, GX\th 9+1, GX\th 9+9 and GX\th 13+1 forming a class of persistently bright atoll sources
(e.g. van der Klis 2006).

The nature of the  banana and island states in atoll sources and of the three Z-track source states
have not been clear and this is central to the basic understanding of LMXB. While there have been 
extensive variability studies (e.g. van der Klis 2006), there have been relatively few spectral 
studies aimed at understanding the island and banana states. Spectral 
investigations may be expected to provide an explanation as revealing directly changes in the non-thermal 
and thermal emission components during changes between states, but there has been a lack of agreement on 
spectral models. Recently, application of a particular approach to the Z-track 
sources (Sect. 1.1) has provided a convincing explanation of those sources and this approach is applied 
in the present paper to the atoll sources.

A further difference between atoll and Z-track sources is that X-ray bursting takes place in 
atoll sources (Grindlay et al. 1976; Lewin et al. 1995), especially at luminosities below 
$2\times 10^{37}$ \erg with large increases of luminosity. These were shown to
consist of unstable nuclear burning of accreted material on the neutron star (e.g. Woosley \& Taam 1976). 
The Z-track sources do not in general display bursting, but strong flares lasting several 
thousand seconds are seen constituting the Flaring Branch (Barnard et al. 2003; Church et al. 2006).

The spectra of LMXB are dominated by non-thermal emission, clearly Comptonized in nature and 
there is also a thermal component 
\begin{figure}                                                                      
\begin{center}         
\includegraphics[width=40mm,height=40mm,angle=270]{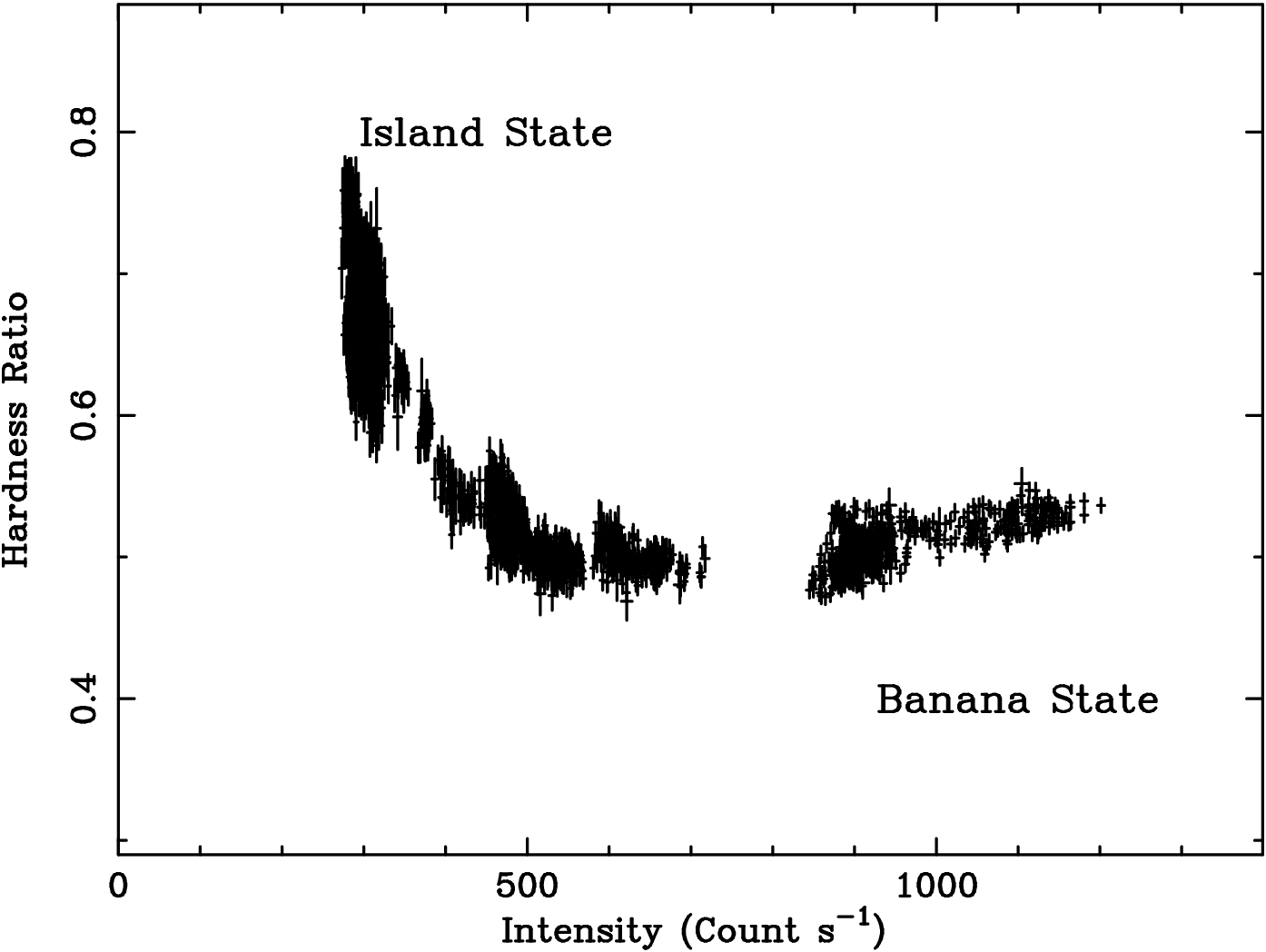}
\includegraphics[width=40mm,height=40mm,angle=270]{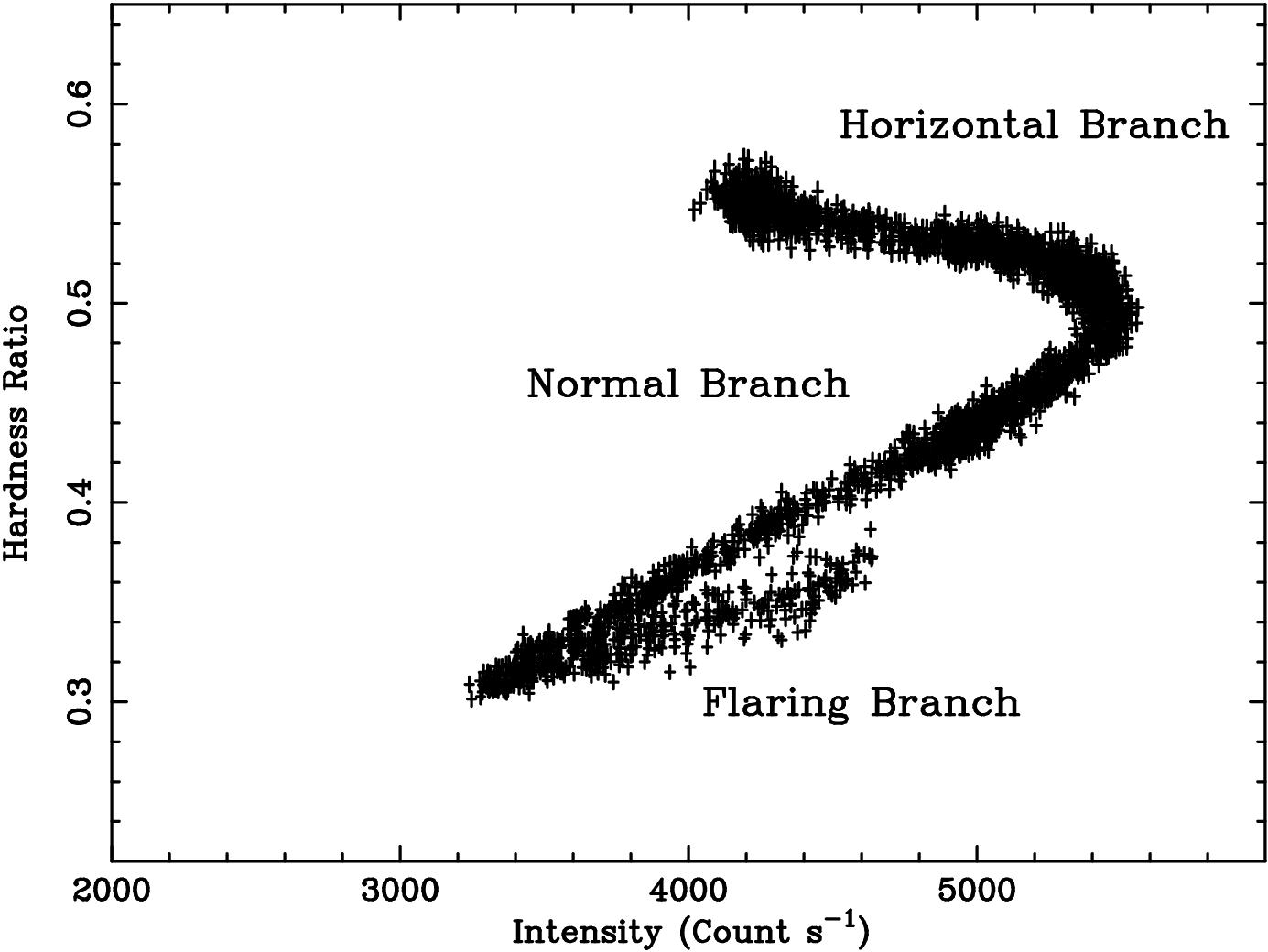}
\caption{Typical hardness-intensity diagrams: left: an atoll source 4U\th 0614+091 showing the
island and banana states and right: a Z-track source GX\th 340+0 showing the Horizontal, Normal and
Flaring Branches (the {\it RXTE} observations of 1998 October and 1997 September, respectively).
Hardness is defined as the ratio of counts $s^{-1}$ in the bands 7.3 - 18.5 keV and 4.1 - 7.3 keV 
and intensity
in the band 1.9 - 18.5 keV.
}

\label{}
\end{center}
\end{figure}
that is weak in the less luminous sources, i.e. the atoll sources (e.g. White et al. 1986; 
Church \& Ba\l uci\'nska-Church 2001). In the Z-track sources, this becomes stronger at up 
to 50\% of the total luminosity (e.g. Church et al. 2006).
LMXB spectra have been fitted with two very different types of model.
In the ``Eastern model'' (Mitsuda et al. 1989), the non-thermal emission is Comptonized
emission from a hot region close to the neutron star and the thermal emission is from the inner 
accretion disk, i.e. multi-colour disk (MCD) blackbody emission. This approach has been
widely applied. In the form developed by Titarchuk and co-workers, Comptonization takes place in
a transition layer between the inner disk and the neutron star (e.g. Titarchuk et al. 1998).
In the Western model (White et al. 1988)
there is Comptonized emission from the accretion disk corona (ADC) above the disk plus blackbody emission 
from the neutron star. 

In terms of these approaches, the hardening of the spectrum in the island state is due in the Eastern 
model to an increase of electron temperature of the central Comptonizing region (e.g. Falanga et al. 2006) 
and in the Western model by an increase in the cut-off energy in the Comptonized emission from an ADC above 
the disk also meaning an increase of electron temperature of the Comptonizing plasma (e.g. Bloser et al. 
2000). However, the cause of the increased temperature in the island state is not understood, nor is
the relation of the atoll to the Z-track sources. One proposal made by Muno et al. (2002) was that if 
observed over long periods, the tracks of the atoll sources displayed in colour-colour look somewhat 
like a Z-track and that the sources are essentially the same. However, van Straaten et al. (2003) found 
evidence against this idea which has not been generally accepted.

Work on the dipping class of LMXB showed that the Comptonizing ADC to extend over a substantial fraction 
of the accretion disk and the Western model was developed into the ``Extended ADC'' model (Church 
\& Ba\l uci\'nska-Church 1995). In these sources, dips in X-ray intensity take place on every binary 
orbit due to absorption in the outer disk (White \& Swank 1982), but the dominant Comptonized emission 
is removed only slowly showing that the emitter is extended (Church et al. 1997). Dip ingress 
timing provided ADC radial sizes of 10\th 000 to 700\th 000 km depending on luminosity (Church \& 
Ba\l uci\'nska-Church 2004). The Extended ADC model gave very good fits to the complex spectral evolution 
in dipping in many observations (Church et al. 1997; 1998a,b, 2005; Ba\l uci\'nska-Church
et al. 1999, 2000; Smale et al. 2001; Barnard et al. 2001). It could also fit the spectra
of all types of LMXB in a survey made with {\it ASCA} (Church \& Ba\l uci\'nska-Church 2001).

The extended nature of the ADC was demonstrated independently in the {\it Chandra}
grating results of Schulz et al. (2009) for Cyg\th X-2. These revealed a wealth
of emission lines of highly excited states (many H-like ions) such as Ne X, Mg XII, Si XIV,
S XIV, S XVI, Fe XXV and Fe XXVI. The line widths seen as Doppler shifts due to orbital motion
indicated emission in a hot ADC at radial positions between 18\th 000 and 110\th 000 km consistent
with ADC modelling (Jimenez-Garate et al. 2002) and in good agreement with the overall ADC size
from dip ingress timing.

This body of evidence for an extended corona suggested applying the Extended ADC model
to the Z-track sources, to test whether it could provide a reasonable explanation of these. 
It was found that it did offer a rather straightforward and convincing explanation.

\subsection{A model for the Z-track sources}

Application of this model to the Cygnus\th X-2 like group of sources: Cyg\th X-2, GX\th 340+0 and 
GX\th 5-1 resulted in the following model (Church et al. 2006: Paper I; 
Jackson et al. 2009: Paper II; Ba\l uci\'nska-Church 2010: Paper III). On the NB, the X-ray 
intensity increases between the soft apex and hard apex due to increasing mass accretion rate $\dot M$ 
shown by the increasing luminosity. The neutron star blackbody temperature increases by a factor
of two giving a large increase of radiation pressure disrupting the inner disk and launching 
the relativistic jets seen in radio on this part of the Z-track. In the FB there is an
increase of blackbody luminosity suggesting an additional energy source on the neutron star. 
At these luminosities there is a critical value of the mass accretion rate per
unit area of the neutron star $\dot m$ dividing stable from unstable burning of the accreted material.
Measured values of 
$\dot m$ at the soft apex agreed well with this critical value showing that flaring is
unstable nuclear burning on the neutron star.

The work was extended to the Sco\th X-1 like sources (Sco\th X-1, GX\th 17+2 and GX\th 349+2) by Church et al. 
(2012: Paper IV). The behaviour on the NB was similar to that of the Cyg-like sources.
The major difference was the stronger FB, while the HB is often absent. Spectral 
analysis showed the major physical difference was the high neutron star temperature at all positions on the 
Z-track of more than 2 keV. We proposed that non-stop flaring in these sources deposits energy on the
neutron star causing the high temperature. The radiation pressure is always high suggesting
that jets should always be present as seems to be the case in Sco\th X-1 (Hjellming 1990).


\subsection{Approach of the present work}

The Extended ADC model is applied in the form {\sc bb + cpl} where {\sc bb} is simple blackbody 
emission of the neutron star and {\sc cpl} is a cut-off power law representing Comptonization in an extended 
corona. The extended nature of the ADC has a major effect on the Comptonized spectrum due to the
nature of the seed photons which are produced by the accretion disk below the ADC. At large radii there
is a large flux of very soft seed photons with energies of 0.1 keV or less so that
the Comptonized spectrum continues to rise below 1 keV; see Church \& Ba\l uci\'nska-Church 
(2004) for a full discussion. The cut-off power law is
a good representation of such a Comptonized spectrum. The disk blackbody acting as the seed for 
Comptonization is not seen directly. By contrast, in various forms of the Eastern model, with Comptonization
taking place in a compact inner region, the seed photons originate from the neutron star or inner disk 
with $kT \sim$1 keV so that the Comptonized spectrum decreases rapidly below 1 keV (e.g. in the
{\sc comptt} model). This differs very markedly from
that of an extended ADC (see Fig. 4 of Church \& Ba\l uci\'nska-Church (2004)). Thus use of
models such as {\sc comptt} etc is inconsistent with the evidence for an extended ADC.
In our fitting, a simple blackbody is used as substantial evidence for its modification 
has not been found except in a small number of bright X-ray bursts. Theoretically, 
electron scattering in the atmosphere may modify the blackbody, however observational evidence is somewhat 
limited and the effect depends on the electron density which is poorly known (Ba\l uci\'nska-Church et al. 2001).

Thus in the present work we take the approach that the evidence favours an extended ADC and we test
the hypothesis that the Extended ADC model can provide a physically reasonable explanation of the atoll 
sources; we do not attempt to apply other Comptonization models, since use of models such as {\sc comptt}
is not consistent with the evidence that the ADC is extended.
We find that the approach adopted does produce an explanation of the island and banana states 
and moreover suggests a simple explanation of the relation between the Atoll sources and the Z-track
sources. Essentially we find that the banana state and Z-track sources are dominated by physics
not previously realized: that there is thermal equilibrium between the neutron star and
Comptonizing ADC. Thus we propose a unified model of Low Mass X-ray Binaries in general.
Previously the relation between Atoll and Z-track sources has not been clear. However, a step towards
a unified model was provided by studies of the transient source XTE\th J1701-462 as it decayed from outburst to
quiescence. This showed that the source underwent a transition from Z-track type to Atoll type (Homan et al. 
2010), including a decay from Z-track to upper banana, strongly indicating that the type depends on mass 
accretion rate, and showing that any effects of 
magnetic field strength and inclination were only secondary. However, it was not clear what caused movement 
along the tracks while in the Z-track or Atoll states. This work is further discussed in Sect. 6.1.


We investigate the banana and island states in the atoll sources in two surveys. Firstly, high quality
{\it Rossi X-ray Timing Explorer (RXTE)} data are used in observations of atoll sources spanning a broad 
range of luminosities. The second survey uses {\it BeppoSAX} with energy range from 0.1 to 300 keV, 
especially valuable for investigating the harder X-ray spectra of the lowest luminosity sources.

Studies using the broadband capabilities of {\it RXTE} and {\it BeppoSAX}
were also made recently of the sources 4U\th 1728-34, GX\th 3+1 and 4U\th 1820-30 by Seifina et al. (2011, 2012)
and Titarchuk et al. (2013) as they moved between different 
states, the main result being that the power law index of the Comptonized emission remained 
constant during state changes. 
The work also showed an increase in the electron temperature of the Comptonizing region in the island
state corresponding to the well-known hardening, applying a model
with Comptonization in a transition layer between the inner disk and the neutron star, a form of 
the Eastern model.

\section{The RXTE survey: observations and analysis}

\tabcolsep 0.8 mm                                                                  
\begin{table}   
\begin{center}
\caption{The {\it Rossi-XTE} Observations.}
\begin{minipage}{84mm}
\begin{tabular}{lrlrr}
\hline \hline
source               &obsid      &date         &span        &exposure\\
                                             &&& (ksec)     & (ksec)\\
\noalign{\smallskip\hrule\smallskip}
4U\th 1916-053       &P10109    &1996 May 14 - 23     &785     &122\\
4U\th 1820-30        &P20075    &1997 Mar 21 - 28     &5865     &46\\
4U\th 1728-34        &P20083    &1997 Sep 19 - Oct 1  &1071     &170\\
SLX\th 1735-269      &P80138    &2003 Sep 18 - Oct 6  &1559     &18\\
2S\th 0918-549       &P90416    &2004 Jun 18 - 29     &1025     &76\\
4U\th 1735-44        &P91025    &2006 Oct 28 - Nov 1  &348      &24\\
4U\th 1636-536       &P91024    &2007 Jan 1 - 15      &1280     &213\\
4U\th 1705-44        &P93060    &2007 Sep 3 - Nov 19  &6686     &216\\
\noalign{\smallskip}\hline
\end{tabular}\\
\end{minipage}
\end{center}
\end{table}
It is important to have long observations in which each source varies substantially in intensity and so moves
on both the banana and island tracks, or moves substantially on one of the tracks,
Thus observations spanning several 100 ksec or more as a series of short exposures were used,
as shown in Table 1 which also gives the total on-source exposure. With short
observations there is often no track movement.
Using local software, the HEASARC {\it RXTE} Archive was examined remotely to produce a hardness-intensity 
diagram for each observation to search for atoll source observations with substantial movement in 
hardness-intensity. 
The observations selected are shown in \hbox{Table 1.} A series of 
observations of each source was made, the total observation in each case spanning $\sim$10 days
(except in 4U\th 1735-44). Substantial changes in intensity took place in each source
as seen in Fig. 2 and there were substantial movements in hardness-intensity (Fig. 3). 

Data from both the Proportional Counter Array (PCA, Jahoda et al. 1996) and the high-energy X-ray timing 
experiment (HEXTE) were used. The PCA was in Standard 2 mode with 16\th s resolution and examination of the 
housekeeping data revealed the number of the xenon proportional counter units (PCU) that were reliably 
on during each observation and these were selected. Standard screening criteria  
were applied to select data having an offset between the source and telescope pointing of less than 
0.02$^o$ and elevation above the Earth's limb greater than 10$\degmark$. Data were extracted from the top 
layer of the detector. Analysis was carried out using the {\sc ftools 6.9} 
\begin{figure*}
\begin{center}                                                               
\includegraphics[width=50mm,height=50mm,angle=270]{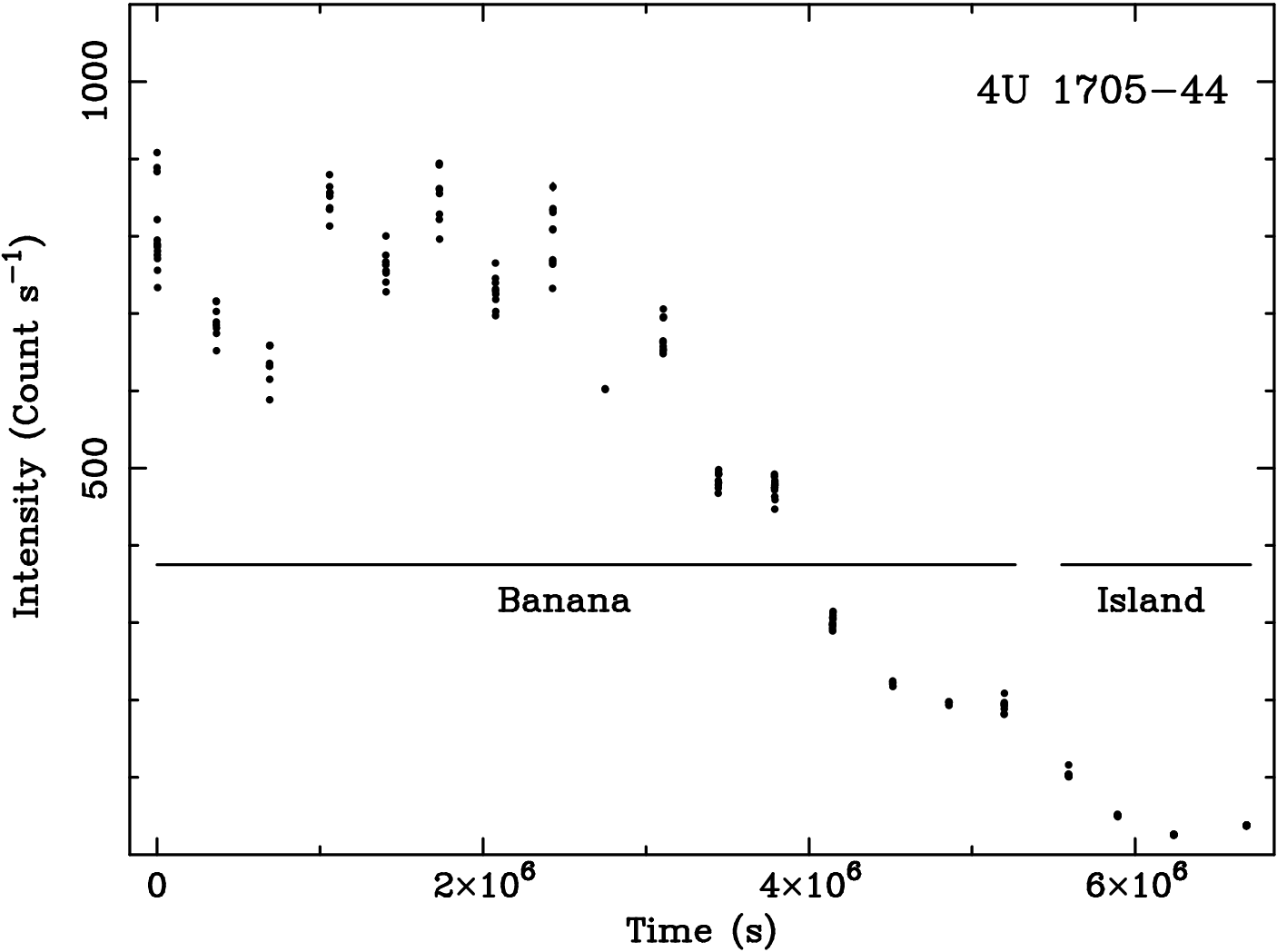}
\includegraphics[width=50mm,height=50mm,angle=270]{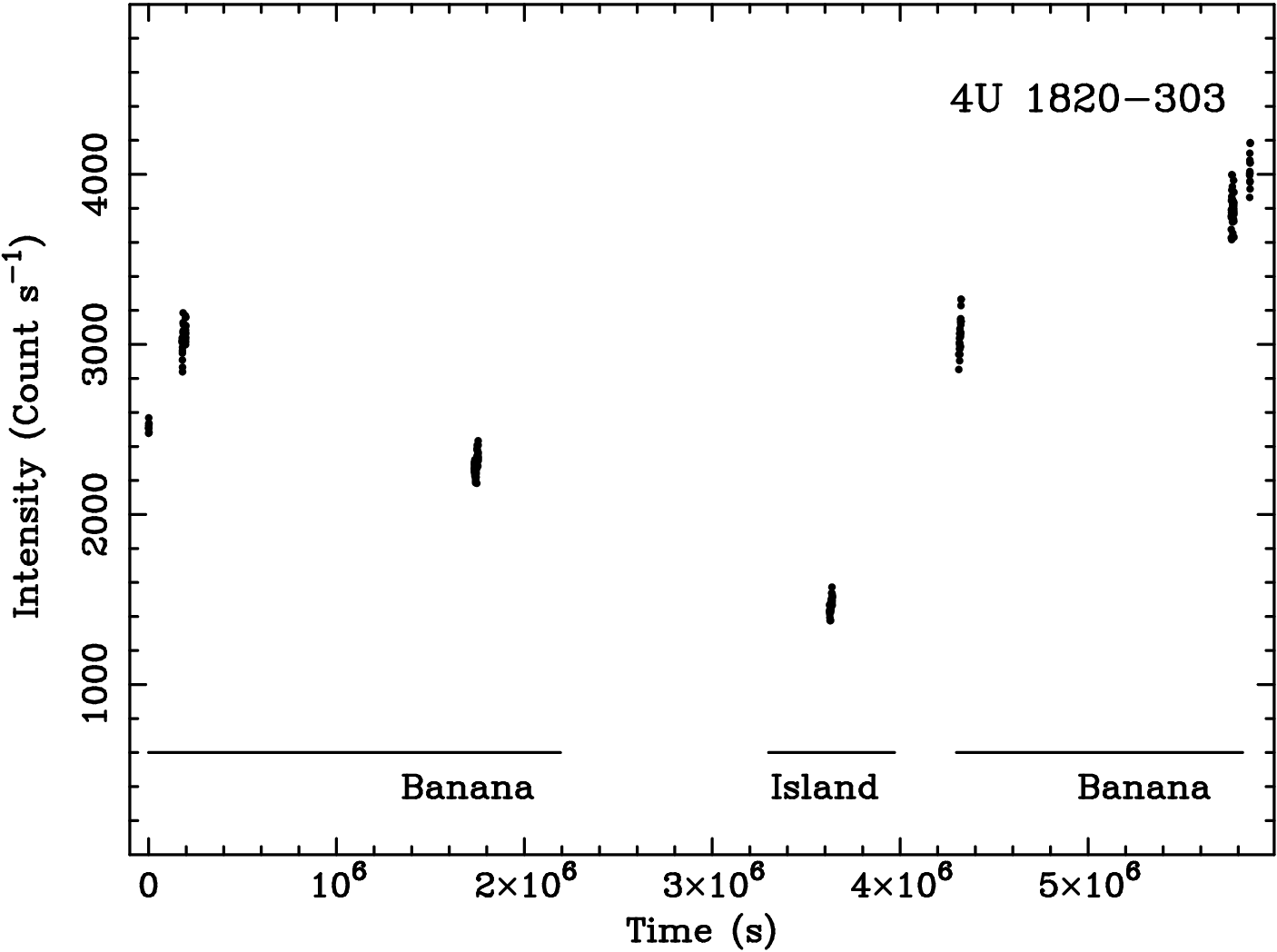}
\includegraphics[width=50mm,height=50mm,angle=270]{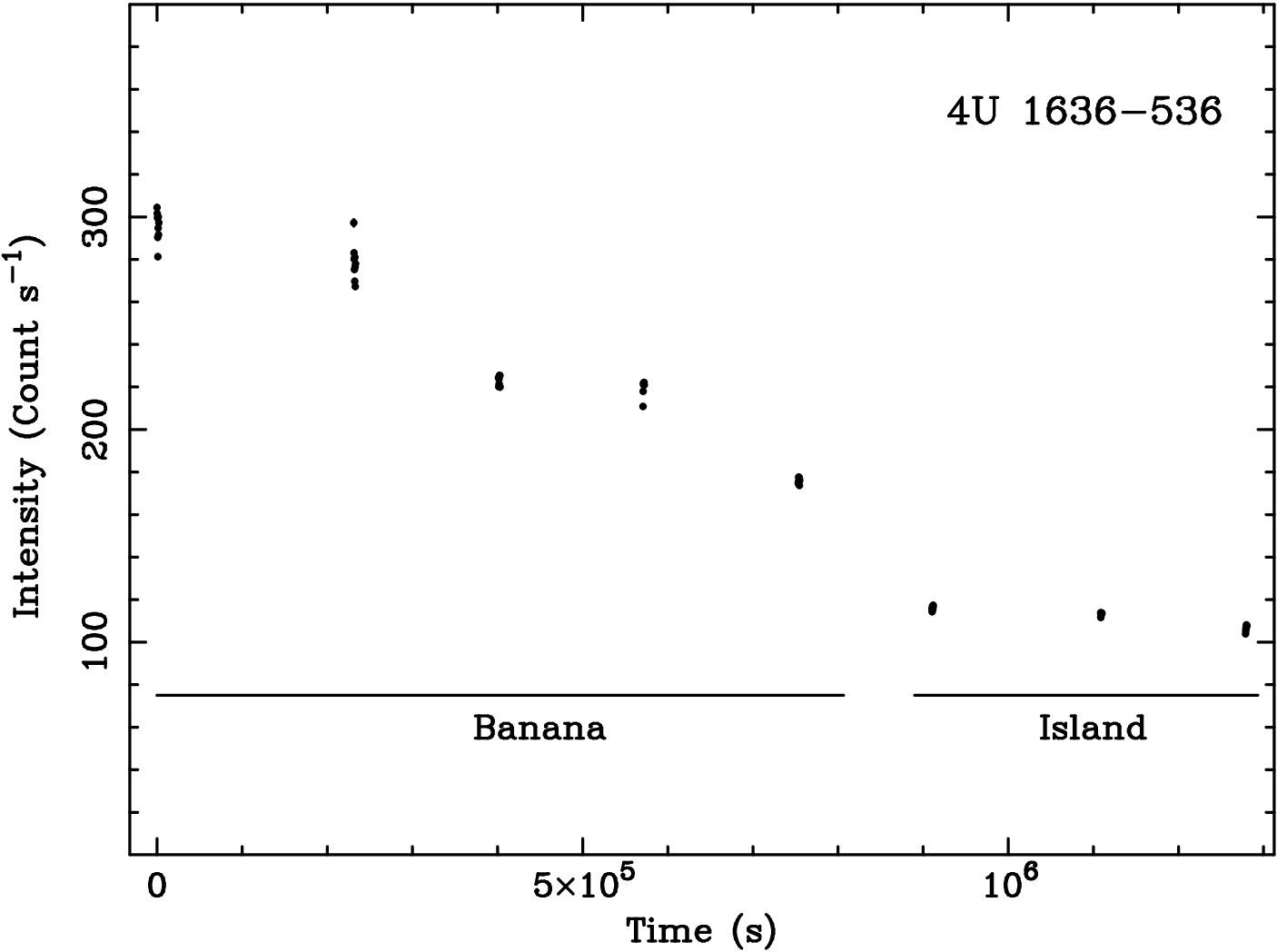}
\includegraphics[width=50mm,height=50mm,angle=270]{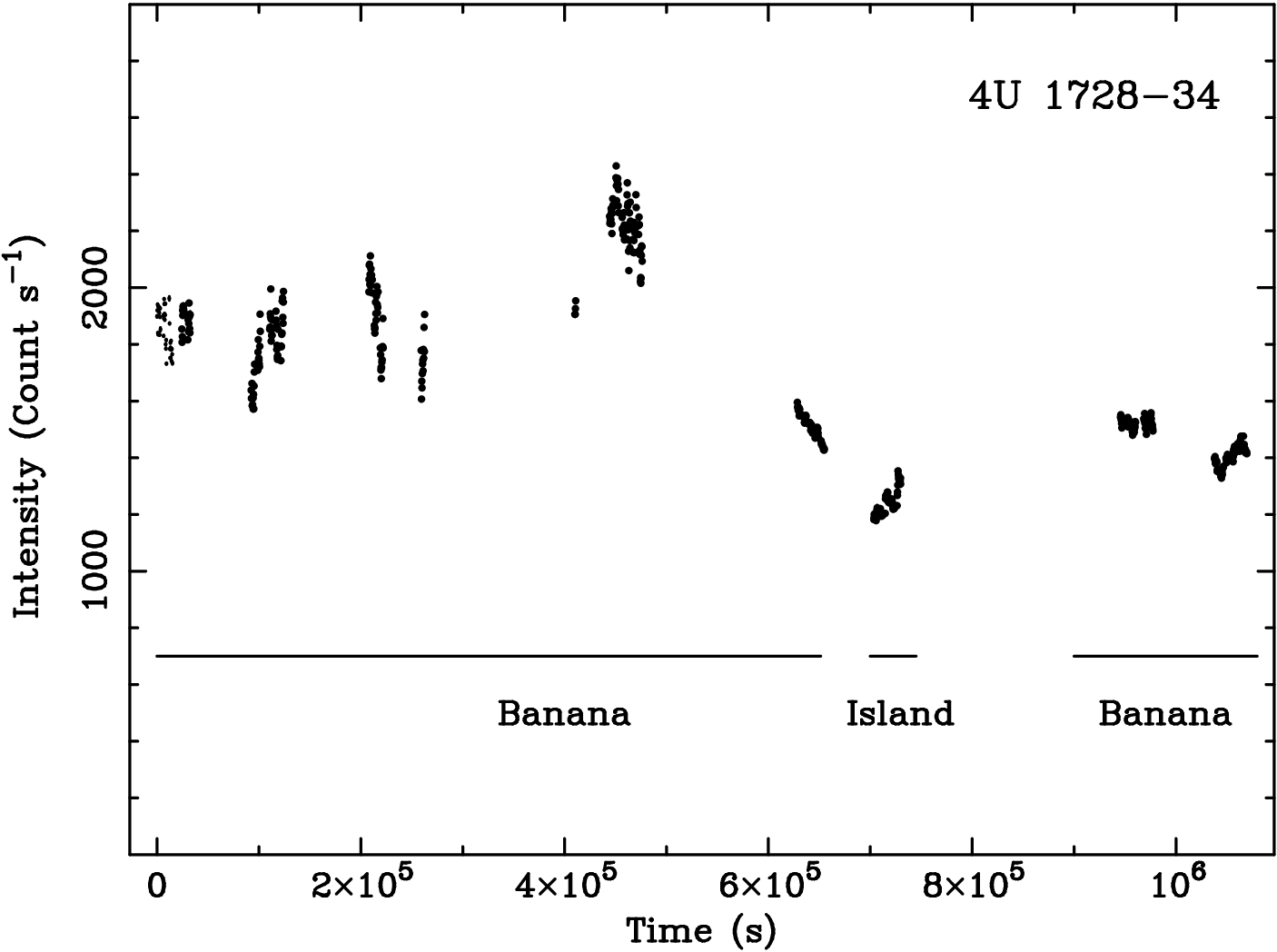}
\includegraphics[width=50mm,height=50mm,angle=270]{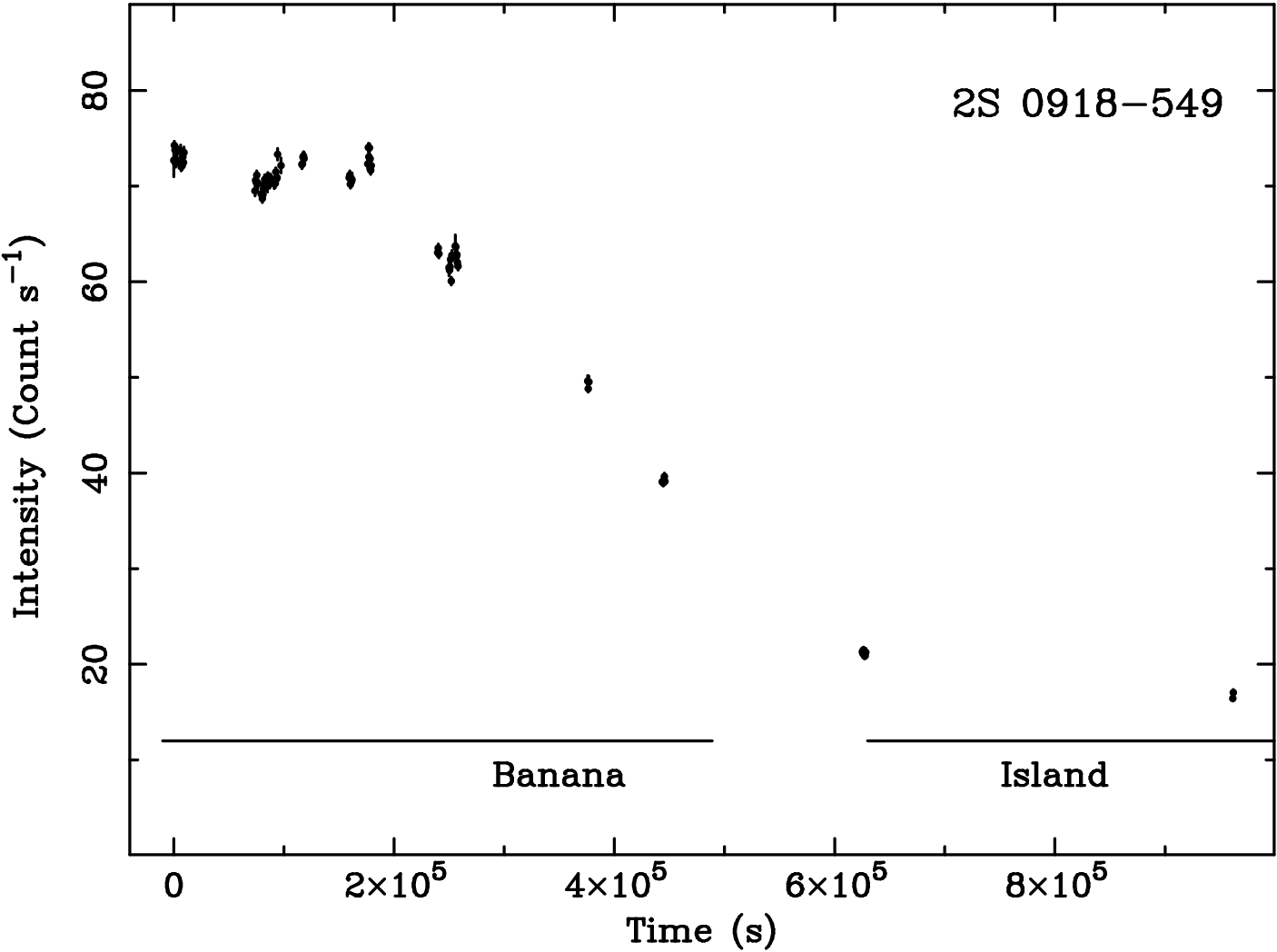}
\includegraphics[width=50mm,height=50mm,angle=270]{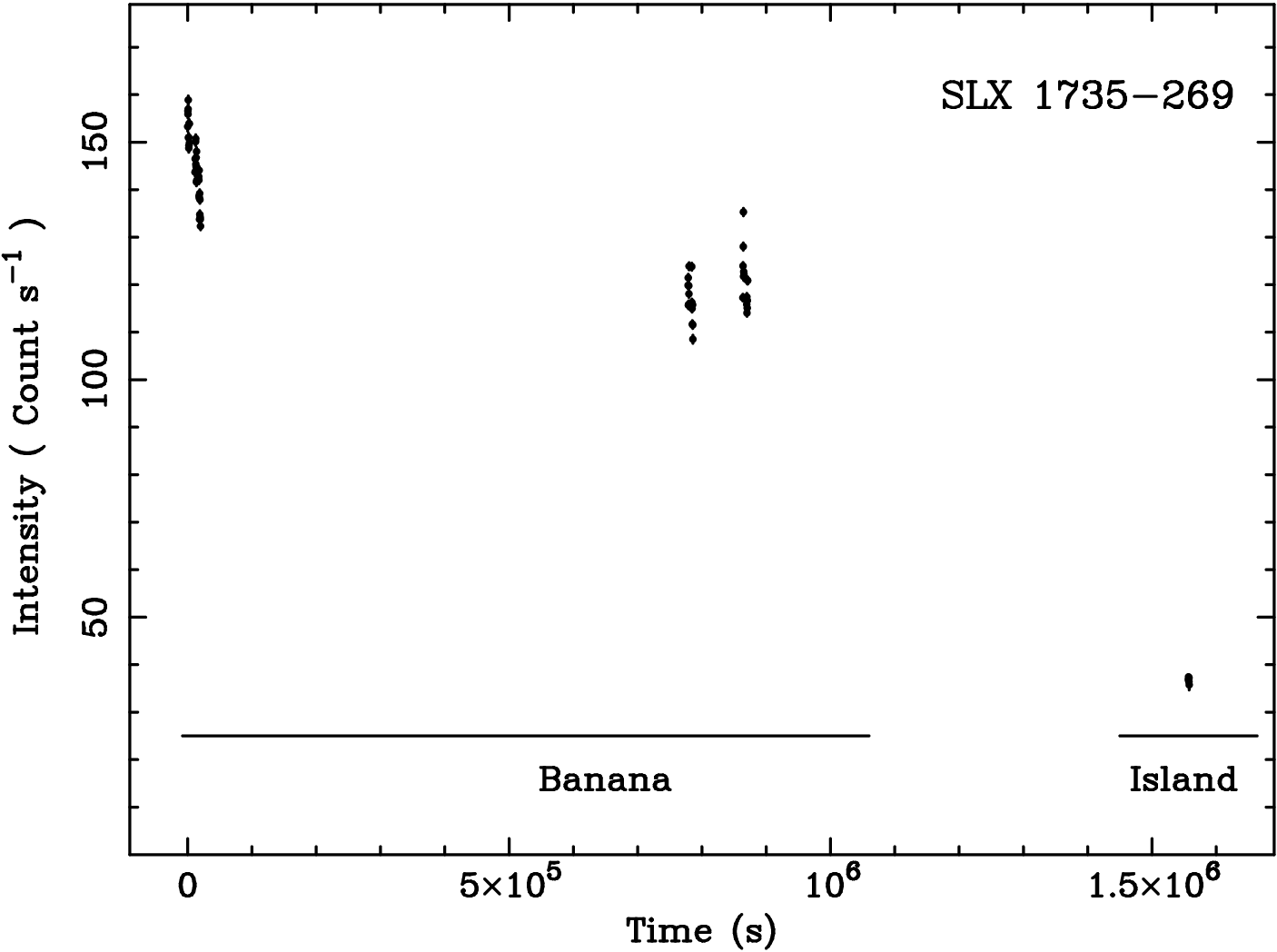}
\includegraphics[width=50mm,height=50mm,angle=270]{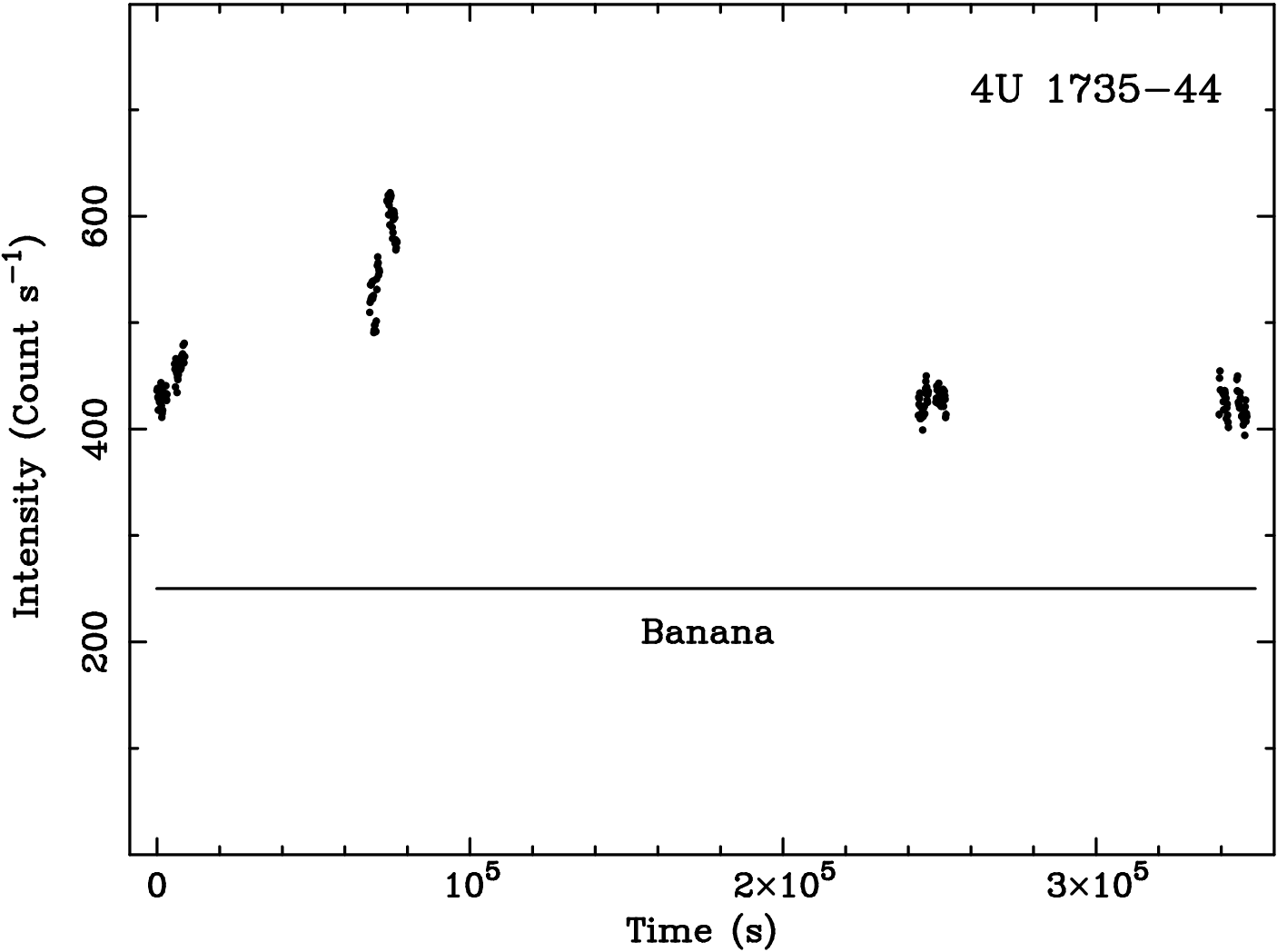}
\includegraphics[width=50mm,height=50mm,angle=270]{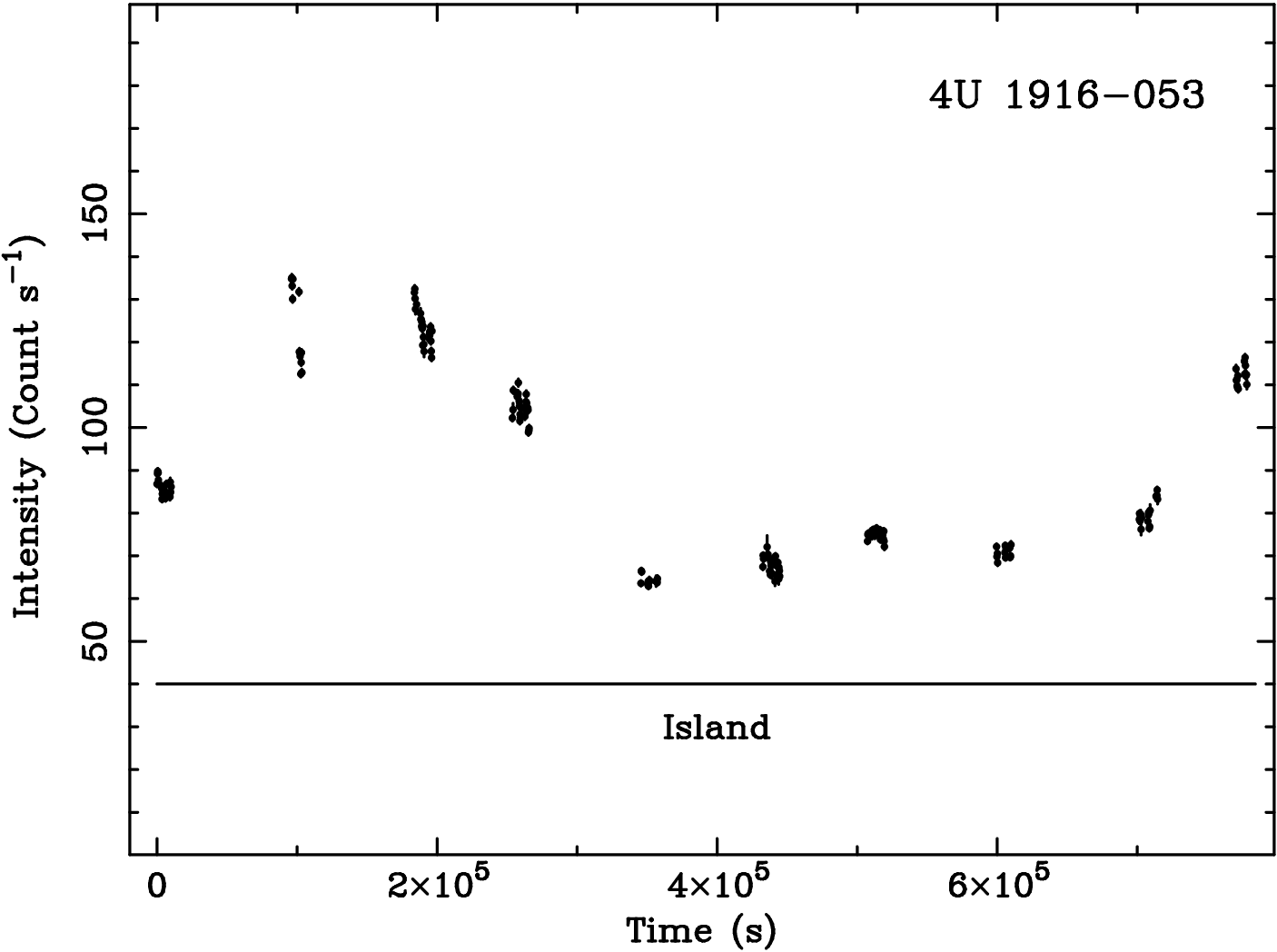}
\caption{Background-subtracted, deadtime-corrected PCA lightcurves of the atoll sources
in the band 1.9 - 18.5 keV. The binning is in general 256 s (see text).}
\label{}
\end{center}
\end{figure*}
package. Total PCA lightcurves 
were extracted to correspond as closely as possible to the band 1.9 -- 18.5 keV for all observations by
selecting appropriate channels. Lightcurves were generated and background files for each PCA data 
file were produced using the facility {\sc pcabackest}, applying the latest background models for background 
subtraction of the lightcurves. Deadtime correction was carried out on both source and background files 
prior to background subtraction and background-subtracted, deadtime-corrected 1.9 -- 18.5 keV lightcurves
are shown in Fig. 2. In most cases a re-binning to 256 s was made to have acceptable Poisson variation except
in 4U\th 1735-44 where 128 s could be used and in 2S\th 0918-459 in which the lower flux required 512 s binning.

Lightcurves were also made in sub-bands of the above range, corresponding to energies 1.9 -- 4.1 keV,
4.1 -- 7.3 and 7.3 -- 18.5 keV. A hardness ratio was defined as the ratio of the intensity in the 
band 7.3 -- 18.5 keV to that in the 4.1 -- 7.3 keV band and the variation of hardness with total
intensity in the band 1.9 -- 18.5 keV is shown for each source in Fig. 3.

In each source an identification of the subsections of data in the lightcurve with banana or island states
was made by selecting each of these sections with the ftool {\sc maketime} and producing a
hardness-intensity plot to find where it lay on the overall track, allowing sections of the lightcurves 
to be labelled as banana or island states.

\section{Results}

\subsection{Lightcurves and hardness-intensity tracks}

The lightcurves for each source shown in Fig. 2; all show a substantial variation in intensity. In most cases 
the source is initially at high intensity in the banana state, the intensity then falling so that
the source makes a transition to the island state. In 4U\th 1705-44 there is a change in intensity of an order 
of magnitude.  4U\th 1735-44 is not seen in the island state, but the light curve exhibits a strong flare 
which spectral analysis shows reaches a luminosity of $\sim 8\times 10^{37}$ erg s$^{-1}$ in the range 
of the GX atoll sources intermediate between the atoll and Z-track sources. 
%
\begin{figure*}
\begin{center}                                                               
\includegraphics[width=50mm,height=50mm,angle=270]{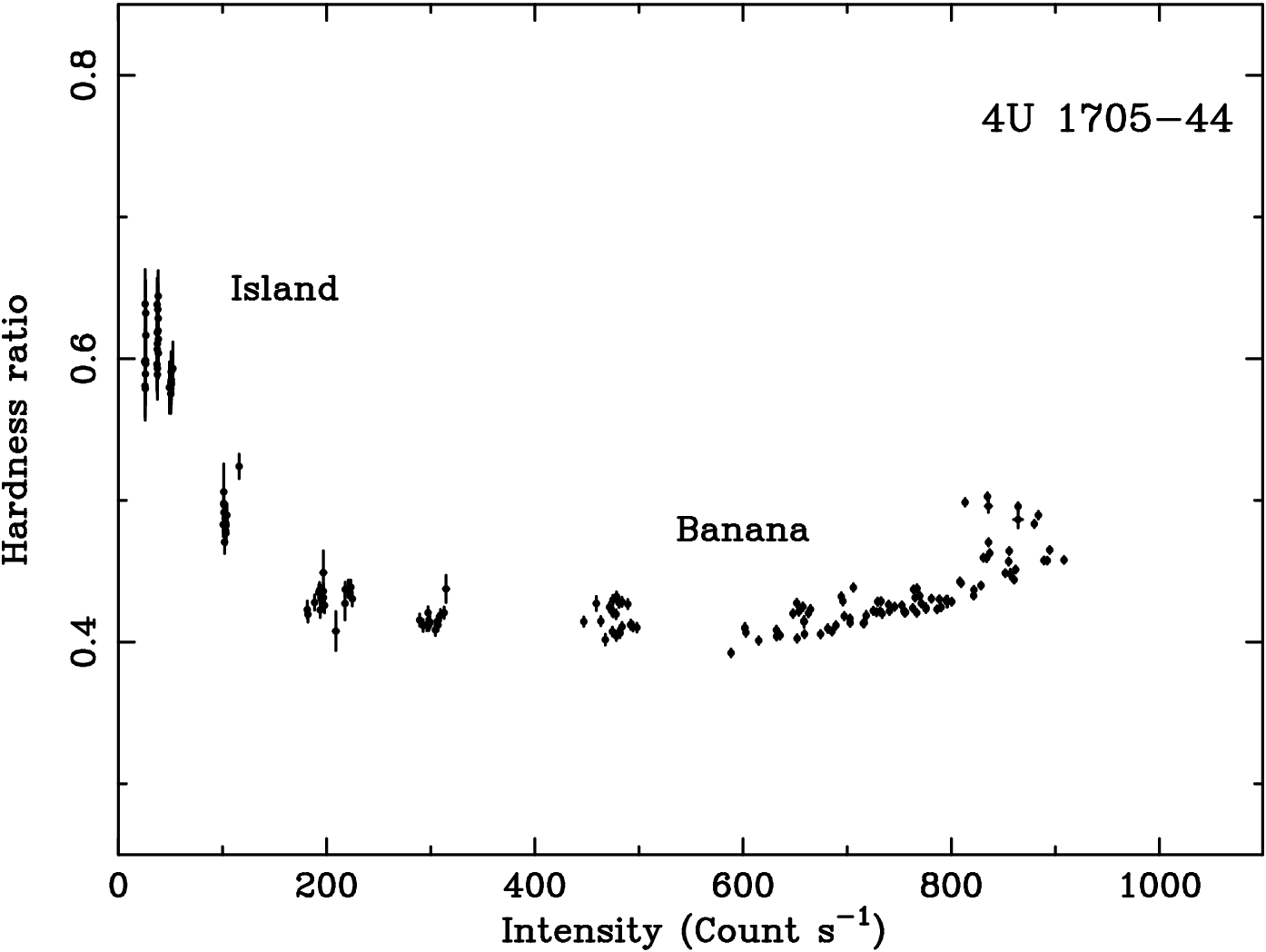}
\includegraphics[width=50mm,height=50mm,angle=270]{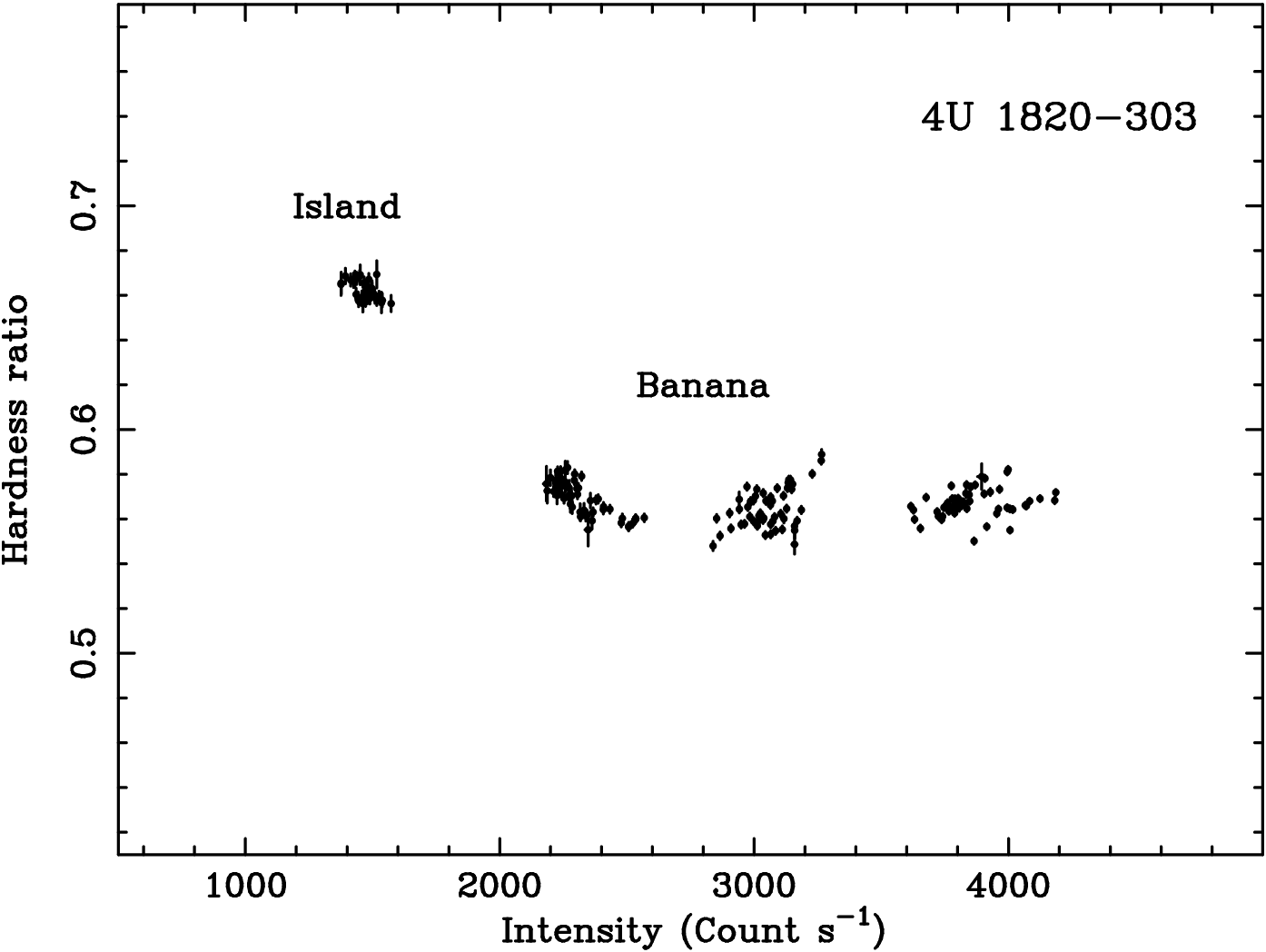}
\includegraphics[width=50mm,height=50mm,angle=270]{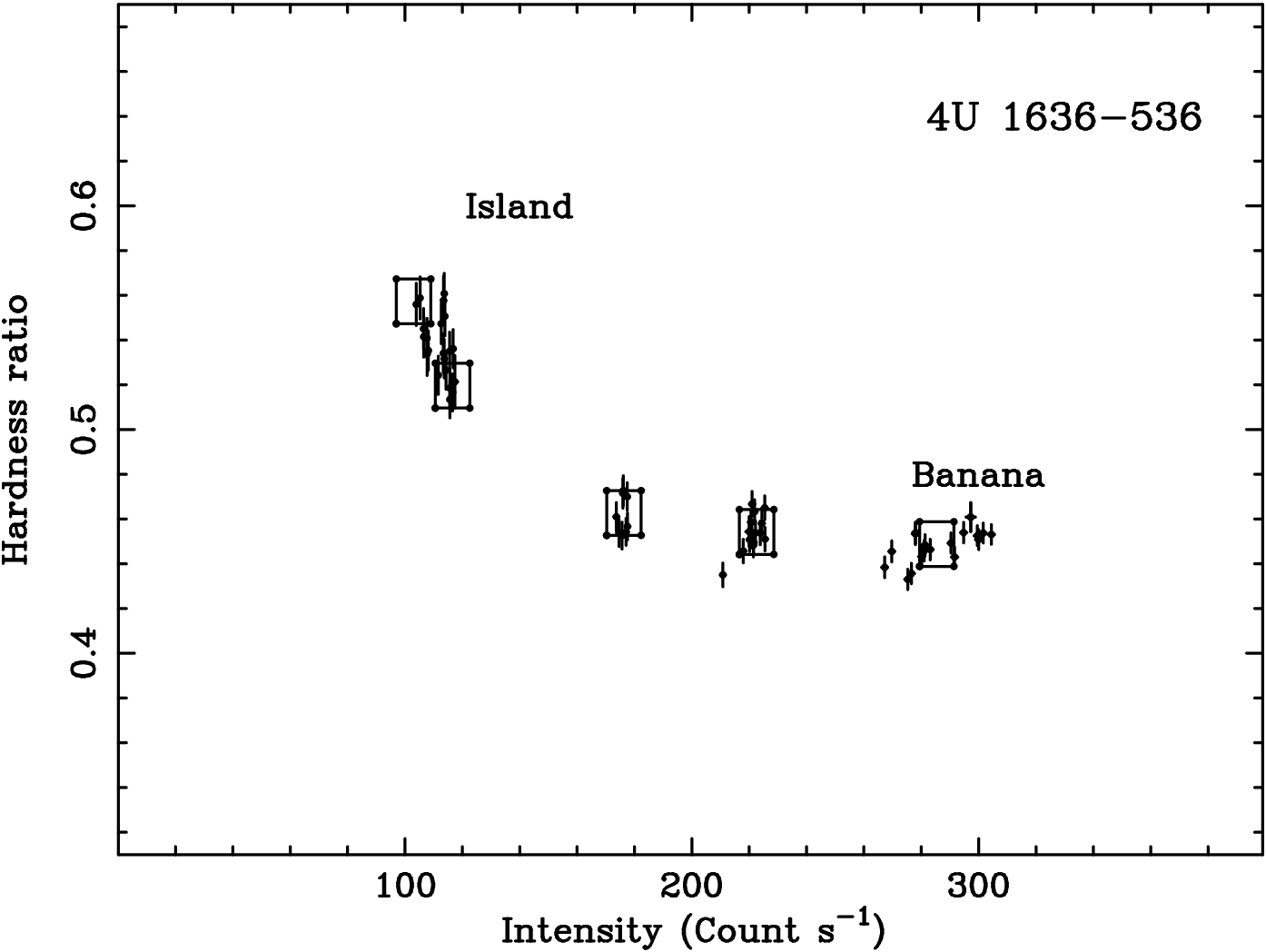}
\includegraphics[width=50mm,height=50mm,angle=270]{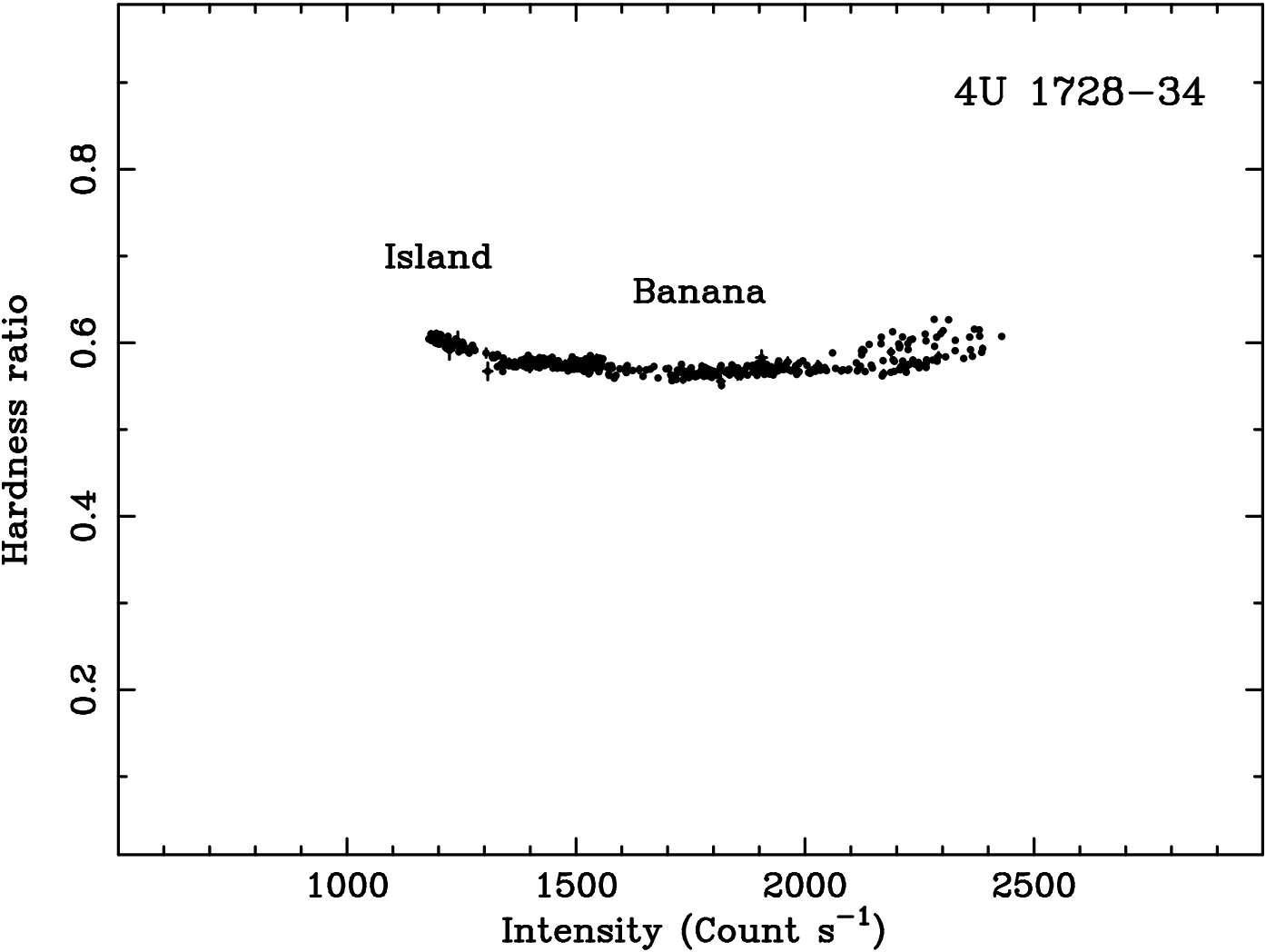}
\includegraphics[width=50mm,height=50mm,angle=270]{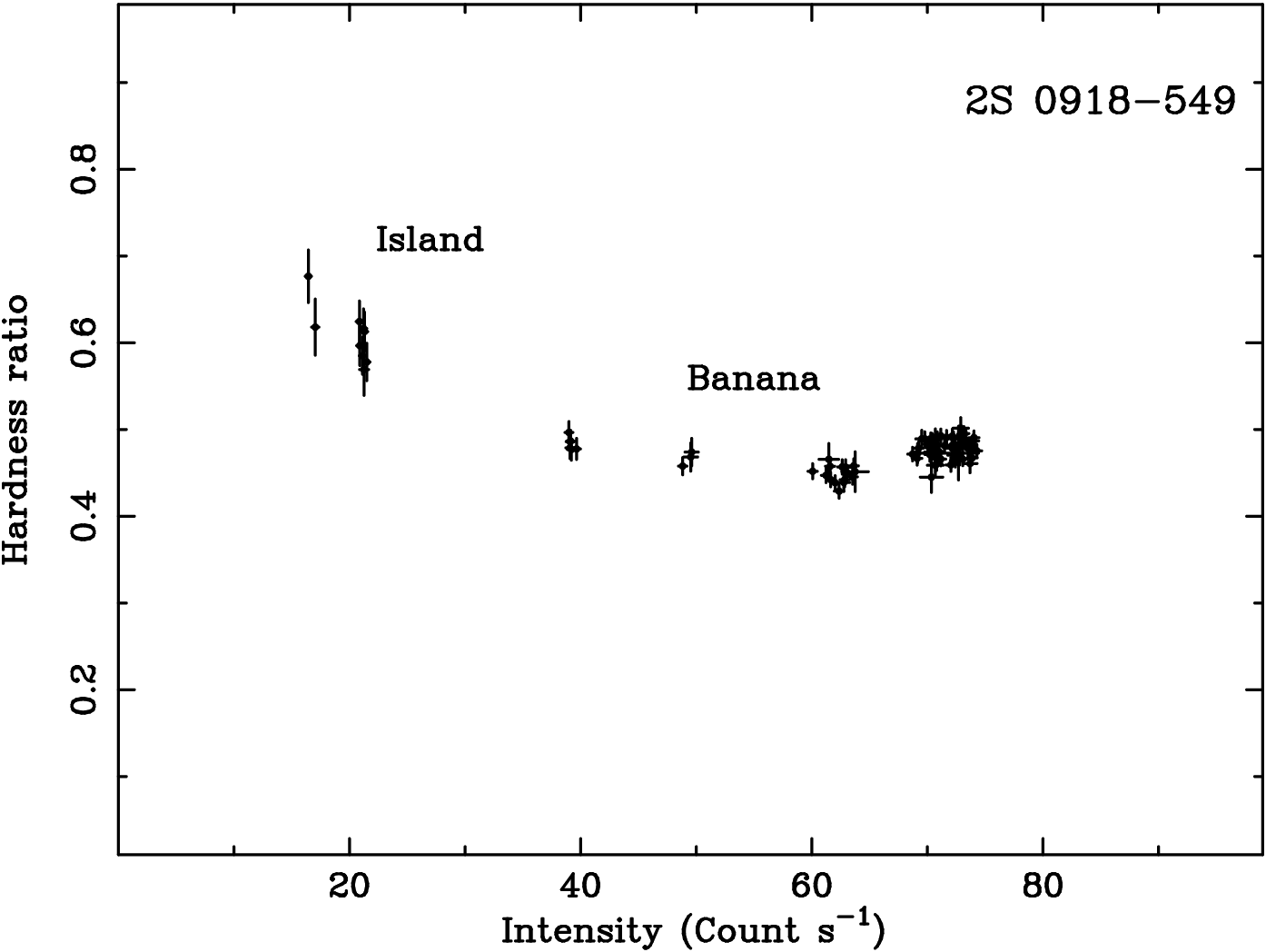}
\includegraphics[width=50mm,height=50mm,angle=270]{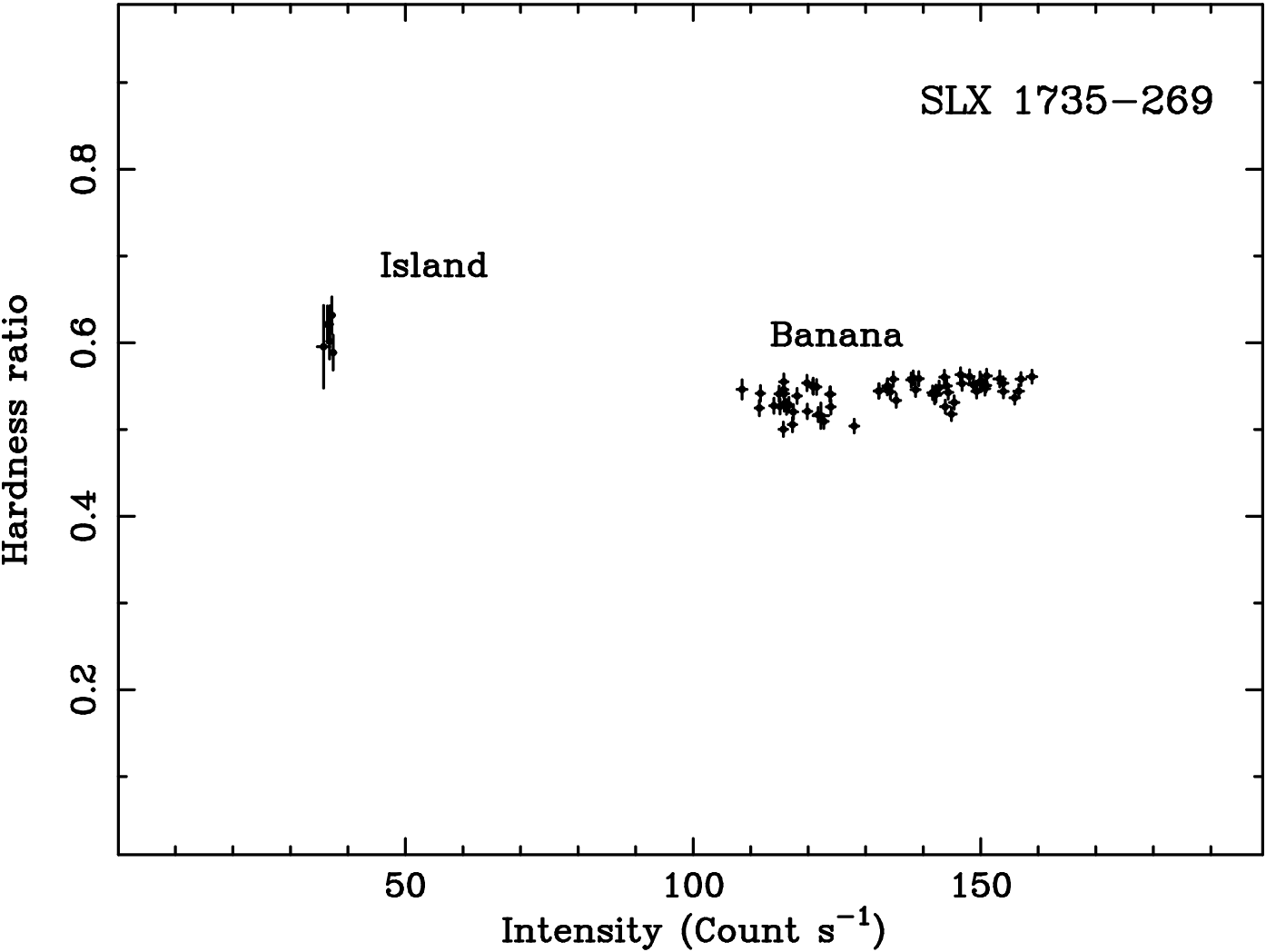}
\includegraphics[width=50mm,height=50mm,angle=270]{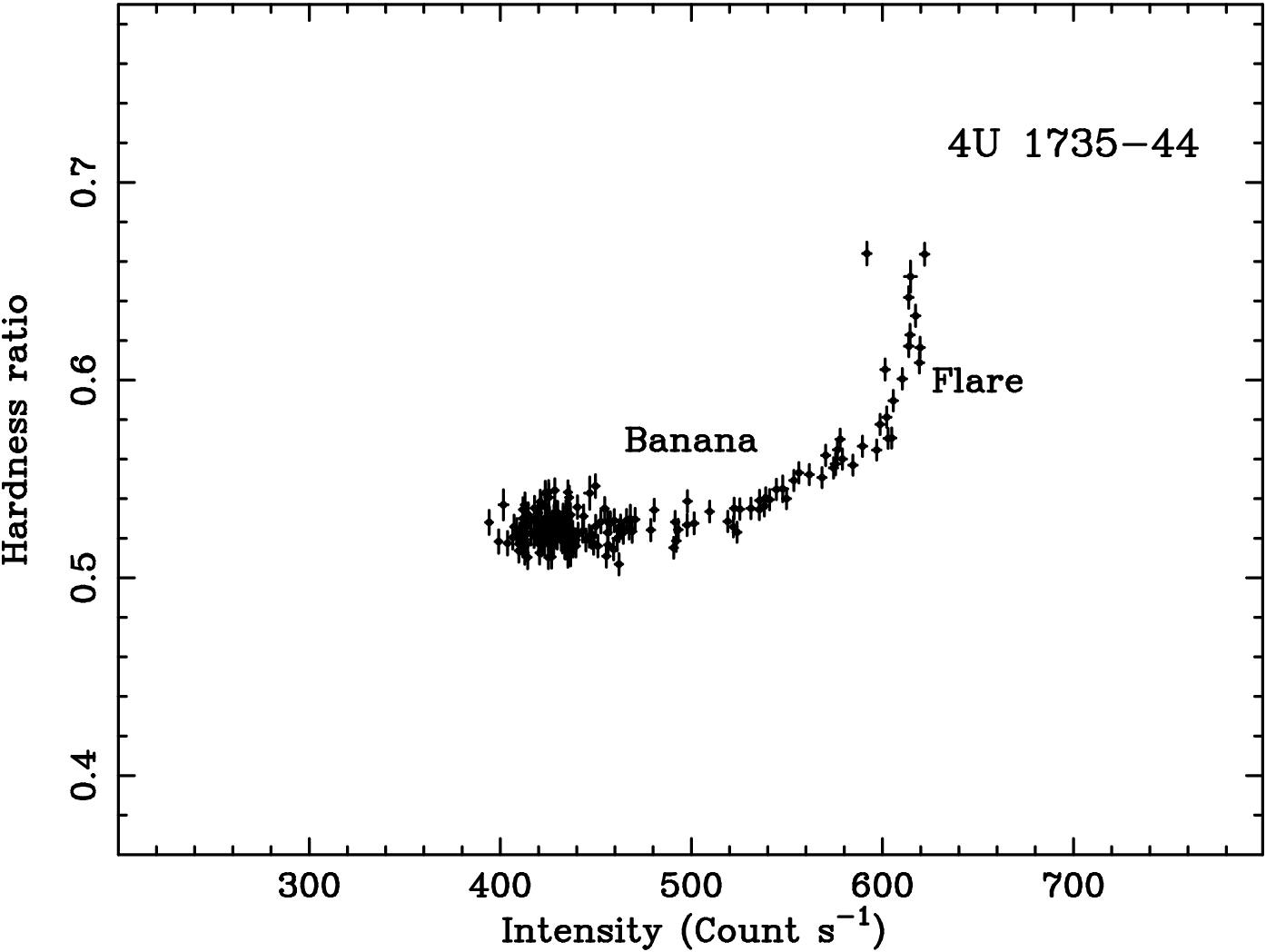}
\includegraphics[width=50mm,height=50mm,angle=270]{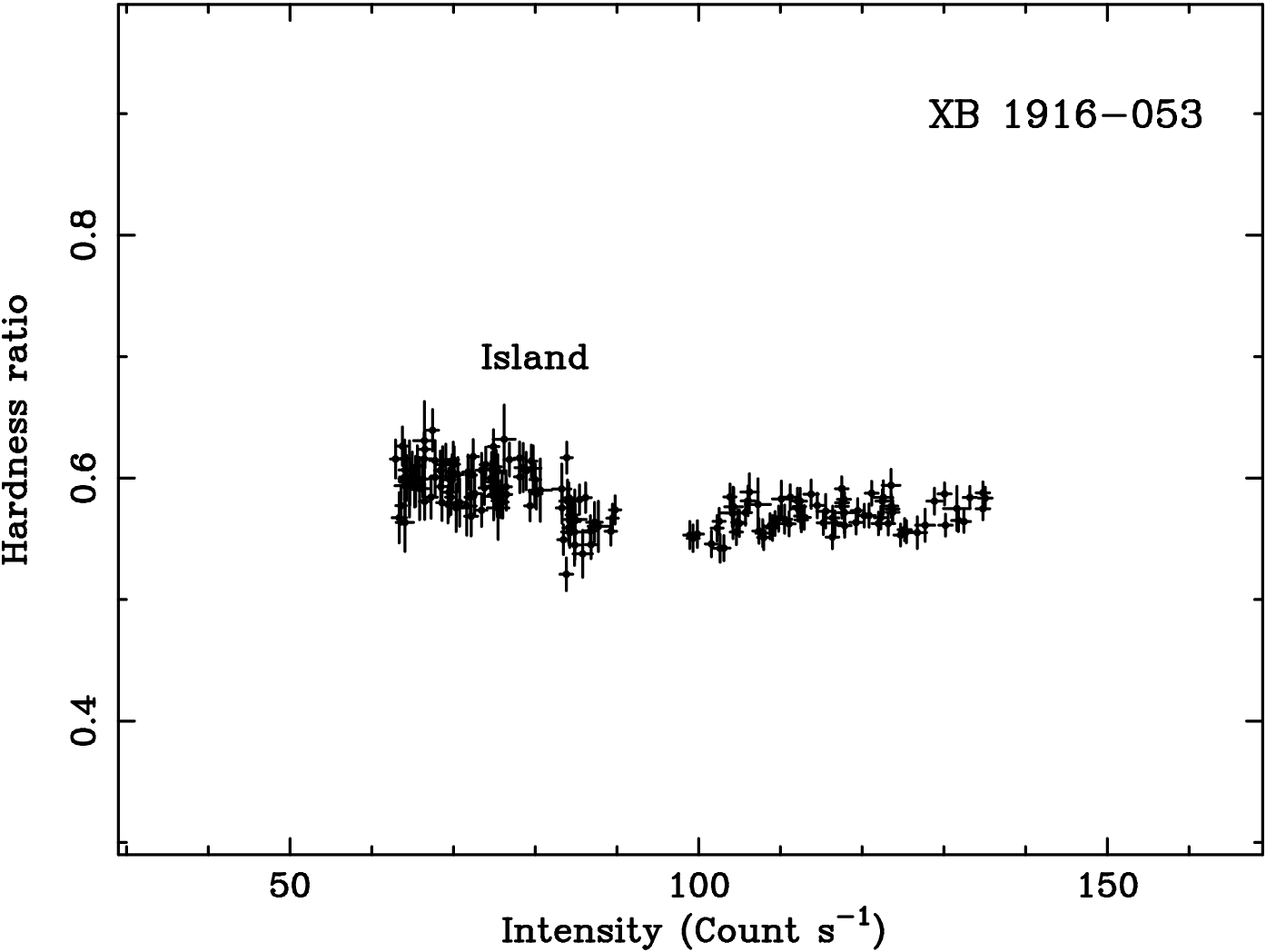}
\caption{Tracks of the survey sources in hardness-intensity from the PCA. Hardness is the ratio of 
counts $s^{-1}$ in the bands 7.3 - 18.5 keV and 4.1 - 7.3 keV and intensity in the total band 1.9 - 18.5 keV.
}
\label{}
\end{center}
\end{figure*}
%
The source is useful in the survey as spanning the higher range of atoll source luminosities.

The corresponding variations in hardness-intensity are shown in Fig. 3 with the tracks labelled as banana or
island states. It can be seen that the initial high intensities in most cases correspond to the rather flat
hardness of the banana state while the low intensity data show the characteristic hardening of the island state.
In 4U\th 1728-34, up-curving in the hardness can be seen at low intensities but the island state
is not reached. Clearly, provided the data span a sufficient range of intensities, as in Fig. 3,
the identification of states is generally quite unambiguous.

\subsection{Spectral analysis}

Spectra were extracted at a series of positions along the tracks executed in each source by defining a box
in hardness-intensity and accumulating a spectrum using Good Time Intervals (GTI) corresponding to data inside
the box. In Fig. 3 we show the boxes used in the case of 4U\th 1636-536; in other cases the boxes are
omitted for clarity. 
\begin{figure*}
\begin{center}                                                               
\includegraphics[width=44mm,height=70mm,angle=270]{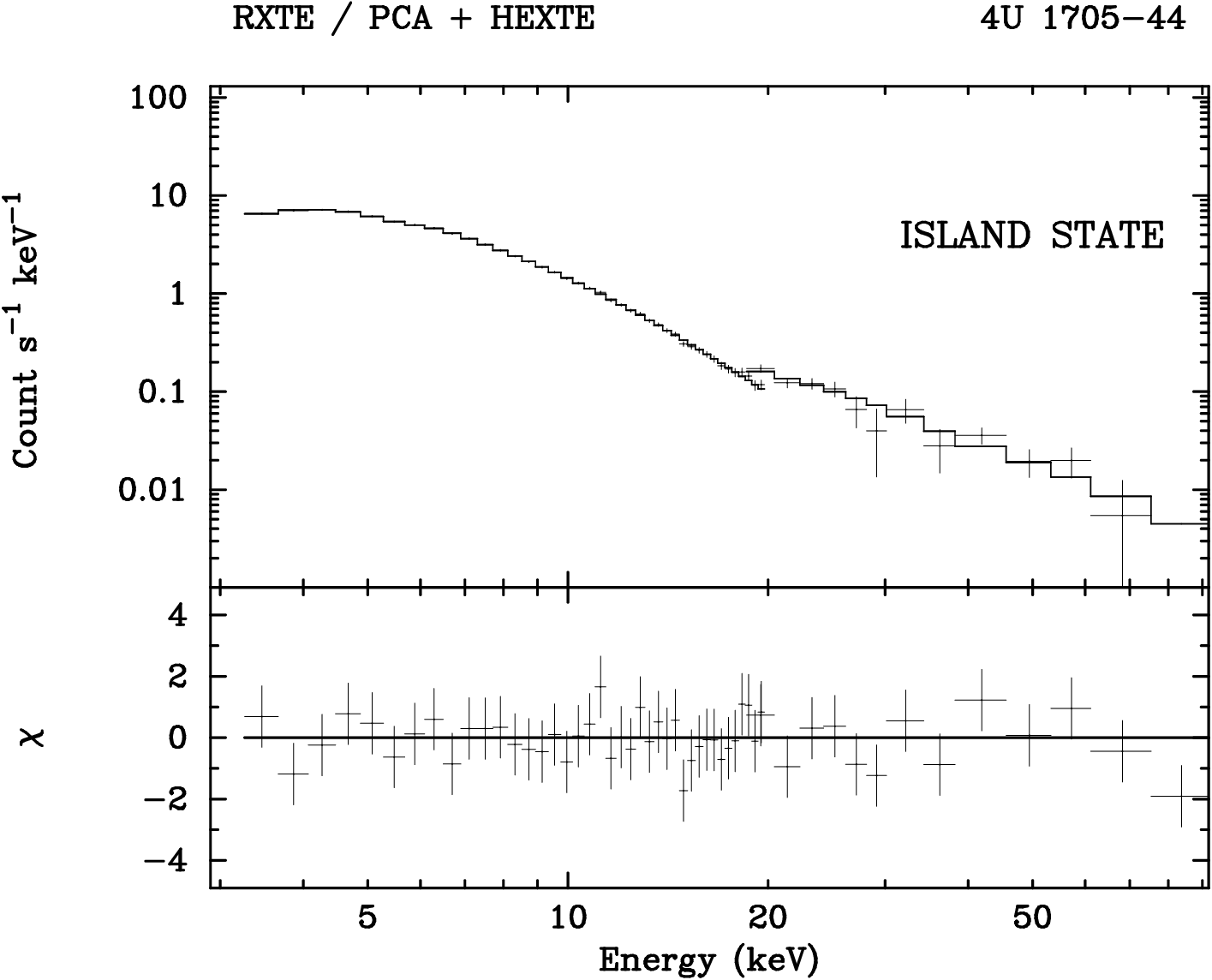}
\includegraphics[width=44mm,height=70mm,angle=270]{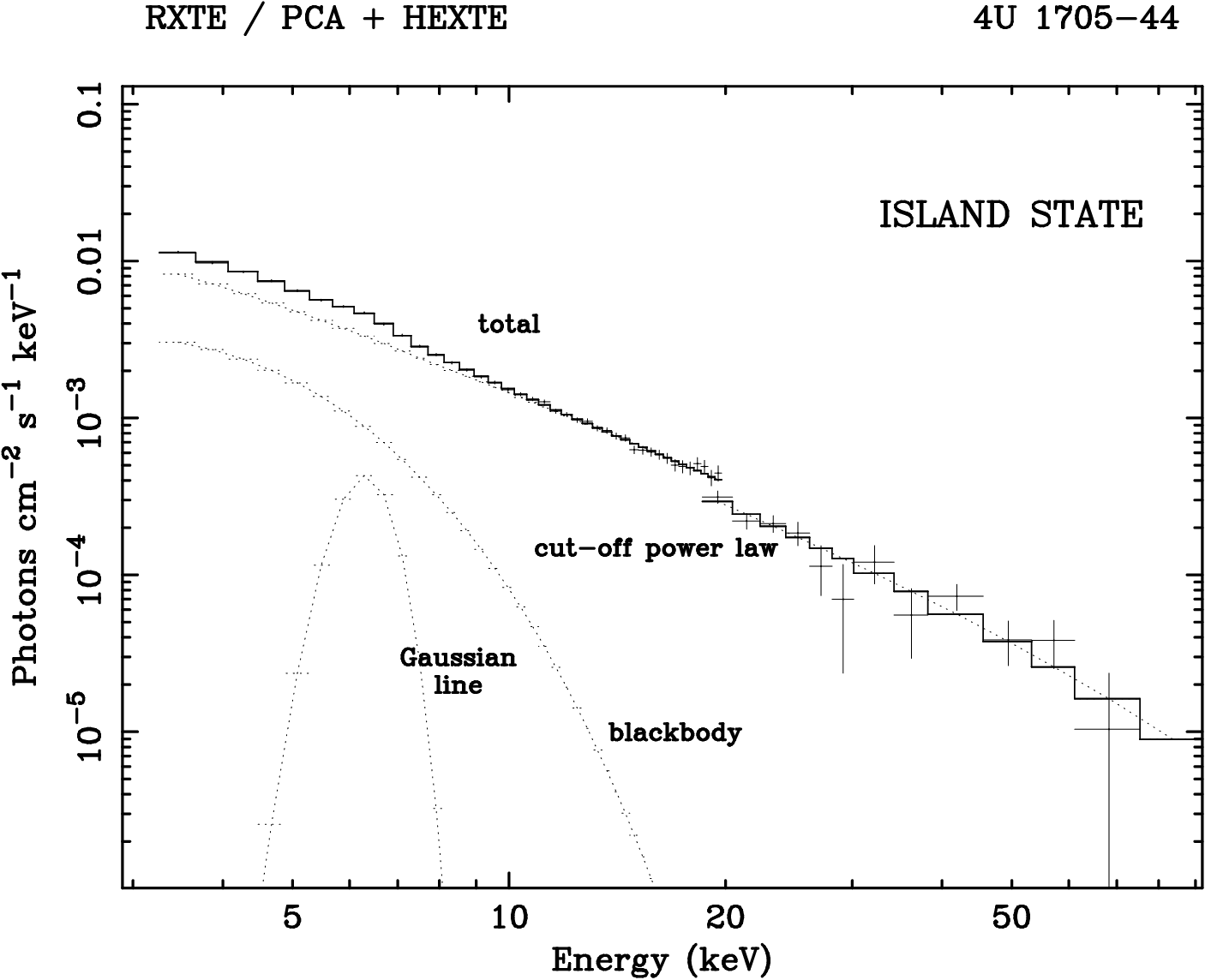}
\includegraphics[width=44mm,height=70mm,angle=270]{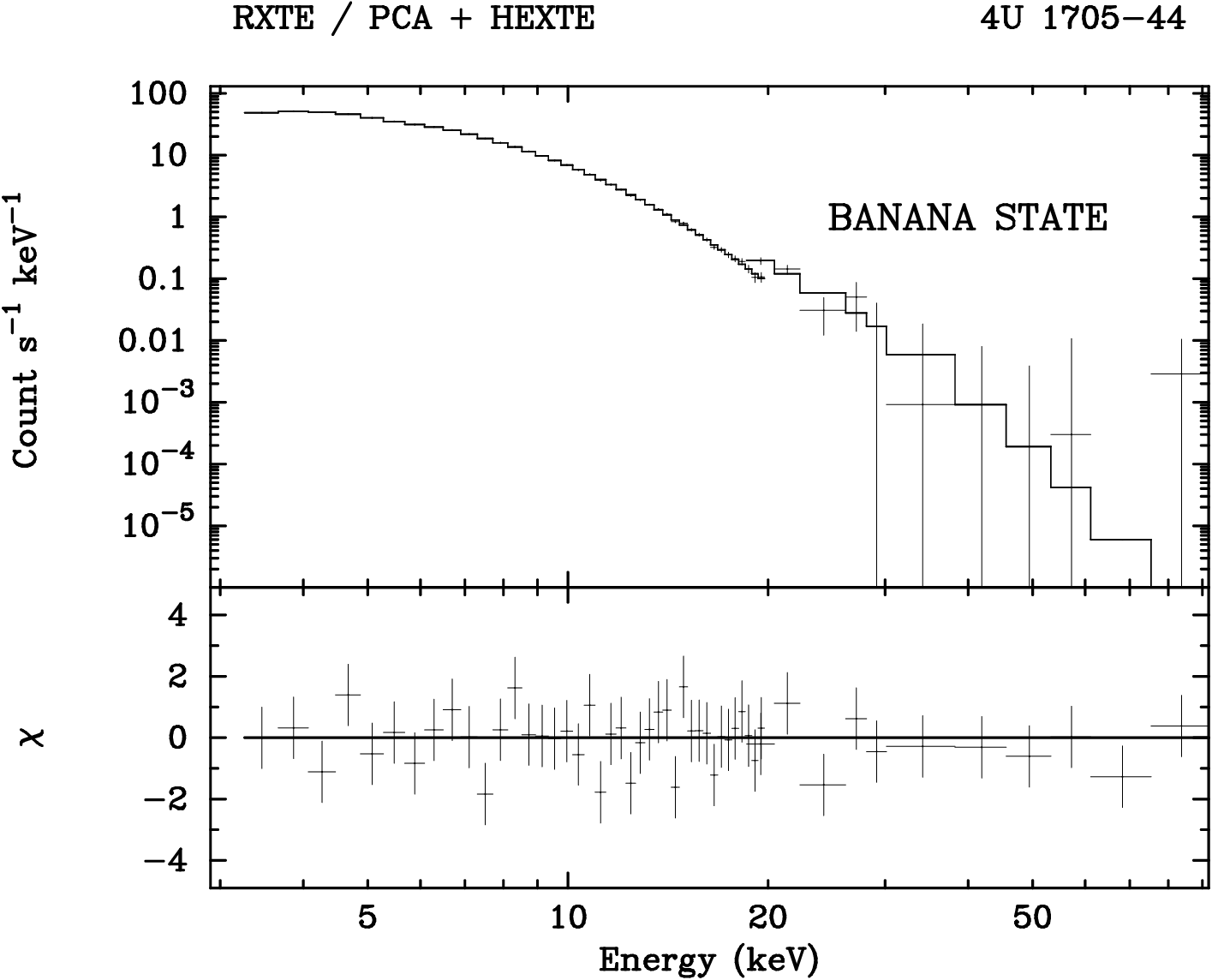}
\includegraphics[width=44mm,height=70mm,angle=270]{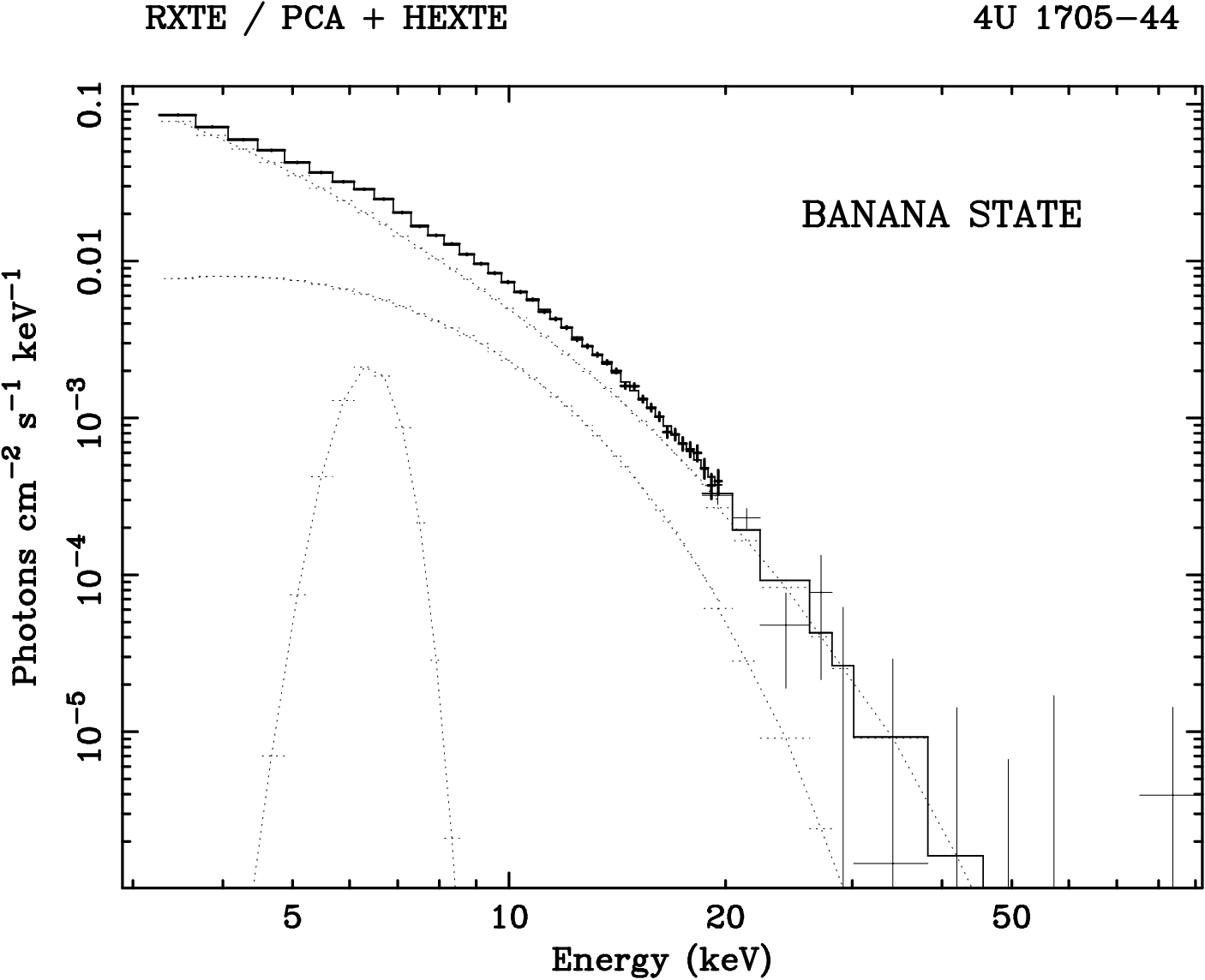}
\caption{Spectral changes between the banana and island state in 4U\th 1705-44: upper panel: the hard spectrum of
the island state (left: folded spectrum; right: unfolded spectrum); lower panel: the soft spectrum of the banana
state (left: folded spectrum; right: unfolded spectrum).
}
\label{}
\end{center}
\end{figure*}
In this source the boxes had a width in intensity of 10 count s$^{-1}$ and a hardness 
range such as 0.50 - 0.52.

Background spectra were produced for each selection and source and background spectra deadtime-corrected
using local software. Channels were re-grouped to a minimum count to allow use of the $\chi^2$
statistic. A systematic error of 0.5\% was added to each channel. Response files were generated using {\sc pcarsp}.

HEXTE source and background spectra were extracted using the same GTI files as for the PCA and deadtime corrected
using {\sc hxtlcurv}. A standard auxiliary response file (arf) and response matrix file (rmf) were used.
The rmf file was re-binned to match the actual number of HEXTE channels using {\sc rddescr}
and {\sc rbnrmf}. No systematic errors were added.
The source + background spectra were compared with the background spectra for both the
PCA and HEXTE, and spectral fitting carried out only up to the energy where these became equal, typically
22 keV in the PCA and 40 keV in HEXTE.

PCA and HEXTE spectra were fitted simultaneously using the extended ADC model allowing a variable normalization 
between the instruments with parameter values chained between the PCA and 
%
%
HEXTE. In cases where the residuals revealed an excess at $\sim$6 keV, a Gaussian line was added to the 
model. Adding a line when there is no indication of an excess in the residuals will distort the 
continuum fitting. A line was seen in all sources except 2S\th 0918-549 and 4U\th 1916-053. In these
sources fitting indicated the equivalent width of any iron line to be less than 5 eV.
Spectral fitting results are shown in Table 2. Although the column density $N_{\rm H}$ can 
often be well-constrained from the part of the
PCA spectrum between 3 and $\sim$15 keV, the lower energy limit of the PCA, usually set in spectral fitting 
at $\sim$3 keV, does not allow accurate determination of low values of the column density and in some 
cases this was fixed at the known Galactic value. Thus in 4U\th 1636-536, $N_{\rm H}$ was set at
$0.3\times 10^{22}$ atom cm$^{-2}$ as also done for XB\th 1916-053. In 4U\th 1735-44, 
$2.6\times 10^{22}$ atom cm$^{-2}$ was adopted. However, $N_{\rm H}$ has relatively little effect
on determination of the other parameters as seen from the error values in the Table.

In atoll sources in the banana state the high energy cut-off of the spectrum can be as low as 5 keV so 
that there is a restricted range of energies for determination of the power law photon index $\Gamma$. 
We have taken the approach that we previously adopted widely in LMXB analysis (including Papers I - IV) 
of fixing $\Gamma$ at 1.7, which seemed optimum and is physically reasonable (Shapiro et al. 1976). This 
procedure gave good quality spectral fits in the Z-track sources and testing showed that varying 
$\Gamma$ did not affect the pattern of variation of spectral fit parameters. 
The assumed constant $\Gamma$ is supported by the work of Seifina et al. (2011, 2012) who found $\Gamma$ not to vary 
during state changes in Atoll source remaining close to 2.0. However, it was noted that the value is 
sensitive to the spectral model. In fitting 4U\th 1728-34 (with a different model), di Salvo et al. (2000) 
obtained a value of 1.60$^{+}_{-}$0.25. Allowing for this sensitivity, the value that we apply
is similar to these. In the second survey using very broadband {\it BeppoSAX} data the values 
may be found by spectral fitting.

Fig. 4 shows spectra of 4U\th 1705-44 to illustrate the high quality 
and the marked difference in hardness between the banana state (lower panels) and the island state
(upper panels). Luminosities were derived from fit results using source distances taken from a careful
examination of the literature (Galloway et al. 2008; Chevalier \& Ilovaisky 1987; Kaminker et al. 1989; 
Goldwurm et al. 1996; Augusteijn et al. 1998; Christian \& Swank 1997.)


\subsection{The luminosity range}

Fig. 5 shows the two decades of luminosity covered by the survey. It is convenient to do this
by showing the variation of the luminosity of the ADC Comptonized emission $L_{\rm ADC}$ 
as a function of the total \hbox{1 -- 30 keV} luminosity $L_{\rm Tot}$.
Logarithmic scales show more clearly that the sources 
analysed span the full range of atoll source luminosities of 10$^{36}$ - 10$^{38}$ erg s$^{-1}$. There 
are relatively few LMXB with luminosities approaching the low level of $1\times 10^{36}$ 
erg s$^{-1}$ and these include XB\th 1916-053 and 2S\th 0918-549 shown here. In some cases, movement of 
the sources along their tracks in hardness-intensity correspond to substantial changes of luminosity 
especially in 4U\th 1705-44. It can be seen that 4U\th 1735-44, 4U\th 1705-44 and 4U\th 1820-30 have luminosities 
in the GX atoll range, approaching the Eddington limit.  

The Comptonized emission is the dominant emission comprising $\sim$85\% of the total 1 - 30 keV
luminosity and the relationship between $L_{\rm ADC}$ and $L_{\rm Tot}$ is close to proportional, 
the data being parallel to the dotted line of slope unity. As the Comptonized emission 
dominates, it is unlikely that $L_{\rm ADC}$ would increase without an increase of mass accretion 
rate $\dot M$ so that
movement to higher luminosities along the banana track corresponds to increase of $\dot M$. 
%
\begin{figure}
\begin{center}                                                               
\includegraphics[width=66mm,height=66mm,angle=270]{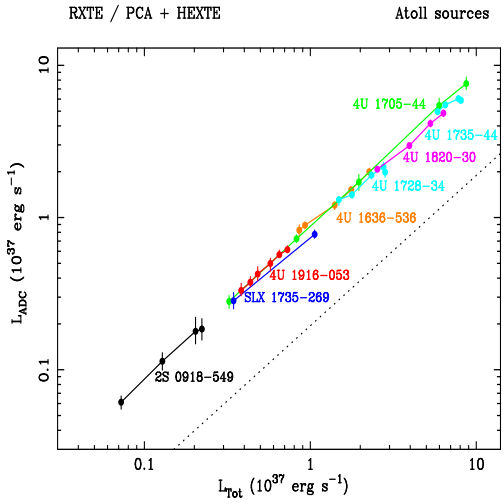}                      
\caption{The luminosity of the ADC Comptonized emission in the band 1 - 30 keV as a function
of the total luminosity in that band, displayed in logarithmic form to show the wide range of
luminosities covered. The dotted line, displaced from the data for clarity, shows the gradient
of a proportional dependence.}
\label{}
\end{center}
\end{figure}
\subsection{The blackbody emission}

The blackbody temperature $kT_{\rm BB}$ was determined directly by spectral fitting, and the blackbody 
radius $R_{\rm BB}$ was derived via $L_{\rm BB}$ = 4$\pi R_{\rm BB}^2 \sigma T_{\rm BB}^4$ where $L_{\rm BB}$ 
is the luminosity of the blackbody and $\sigma$ is \hbox{Stefan's} constant.
In Fig. 6, $kT_{\rm BB}$ (upper panel) and $R_{\rm BB}$ (middle panel) are shown
as a function of $L_{\rm Tot}$. The blackbody temperature varies between 
$\sim$1 and 2 keV, but is relatively constant over the banana state. The blackbody radius at low luminosities
of $\sim 1\times 10^{37}$ erg s$^{-1}$ is of the order of 1 km, indicating that the emission is from a restricted
part of the neutron star, i.e. an equatorial belt in the orbital plane. A similar result was obtained in a previous
{\it ASCA} survey (Church \& Ba\l uci\'nska-Church 2001). 
\tabcolsep 2 mm
\begin{table*}   
\begin{center}
\caption{Results of fitting the Extended ADC model to the {\it RXTE} spectra of the atoll sources.}
\begin{tabular}{lrrrrrrrrrr}
\hline\hline
\noalign{\smallskip}
Source & spectrum & $N_{\rm H}$ & $kT_{\rm BB}$ & norm & $E_{\rm line}$ & EW & $E_{\rm CO}$ & norm & $\chi^2$/d.o.f. \\
\noalign{\smallskip\hrule\smallskip}

2S\th 0918-549 & i1 & 0.3 & $0.64\pm0.08$ & $0.45\pm0.08$ & - & - &  $37_{-10}^{+21}$ & $0.25\pm0.03$ & $ 35/36 $ \\
               & b1 & $0.3$ & $0.91\pm0.06$ & $0.55\pm0.17$ & - & - & $12.1\pm1.8$ & $0.71\pm0.09$ & $ 35/41 $ \\
               & b2 & $0.5_{-0.5}^{+1.0}$ & $1.45\pm0.11$ & $0.9\pm0.3$ & - & - & $8.4\pm0.9$ & $1.4\pm0.3$ & $ 27/40 $ \\
               & b3 & $0.8\pm0.8$ & $1.71\pm0.11$ & $1.3\pm0.2$ & - & - & $8.0\pm0.9$ & $1.5\pm0.2$ & $ 42/42 $ \\
\hline
4U\th 1636-536& i2 & $0.3$ & $1.43\pm0.19$ & $0.7_{-0.7}^{+0.9}$ & $7.1\pm0.3$& $165$ & $15.5\pm1.5$ & $3.6\pm0.3$ & $ 19/30 $ \\
              & i1 & 0.3 & $1.49\pm0.65$ & $0.9\pm0.6$ & $6.6\pm0.3$ & $108$ & $13.0\pm0.4$ & $4.2\pm0.2$ & $ 28/35 $ \\
              & b1 & 0.3 & $2.15\pm0.16$ & $5.1\pm1.1$ & $7.3\pm0.4$ & $92$ & $6.6\pm0.7$ & $8.6\pm0.5$ & $ 36/36$ \\
              & b2 & 0.3 & $2.08\pm0.10$ & $6.5\pm0.7$ & $6.8\pm0.3$ & $62$ & $6.6\pm0.5$ & $10.8\pm0.5$ & $ 32/35$ \\
              & b3 & 0.3 & $1.88\pm0.16$ & $7.2\pm0.9$ & $6.8\pm0.4$ & $114$ & $7.4\pm0.5$ & $13.1\pm0.9$ & $ 26/36$ \\
\hline
4U\th 1705-44 & i2  & $1.8\pm0.7$ & $1.11\pm0.10$ & $1.25\pm0.19$ & $6.31\pm0.18$ & $130$ & $62_{-17}^{+46}$ & $0.87\pm0.08$ & $35/45 $ \\
              & i1  & 2.0 & $1.68\pm0.13$ & $2.9\pm0.4$ & $6.18\pm0.15$ & $115$ & $9.4\pm0.5$ & $4.4\pm0.3$ & $ 35/43 $ \\
              & b1 & $2.1\pm0.6$ & $1.95\pm0.15$ &  $7.0\pm0.6$ & $6.40\pm0.13$ & $111$ & $5.8\pm0.5$ & $14.4\pm1.8$ & $ 35/43 $ \\
              & b2 & $4.7\pm0.8$ & $1.94\pm0.16$ & $14.2\pm1.9$ & $6.42\pm0.16$ & $183$ & $5.3\pm0.3$ & $44\pm4$ & $ 52/42 $ \\
              & b3 & $5.8\pm0.7$ & $2.01\pm0.08$ & $31.3\pm1.8$ & $6.64\pm0.17$ & $184$ & $5.3\pm0.2$ & $59\pm4$ & $ 43/43 $ \\
\hline
4U\th 1728-34 & i1 & $1.9\pm0.4$ & $1.92\pm0.07$ & $6.0\pm0.8$ & $6.54\pm0.08$ & $161$ & $14.8\pm0.9$ & $7.7\pm0.5 $ & $ 38/46 $ \\
              & b1 & $2.4\pm0.3$ & $2.21\pm0.04$ & $13.0\pm0.3$ & $6.61\pm0.10$ & $132$ & $8.2\pm0.3 $ & $11.5\pm0.6 $ & $ 27/48 $ \\
              & b2 & $3.4\pm0.3$ & $2.18\pm0.05$ & $15.4\pm0.4 $ & $6.63\pm0.10$ & $141$ & $7.7\pm0.3 $ & $16.1\pm0.9$ & $ 41/48 $ \\
              & b3 & $3.6\pm0.4$ & $2.10\pm0.06$ & $22.0\pm0.7$ & $6.63\pm0.13$ & $118$ & $7.8\pm0.5 $ & $18.1\pm1.5$ & $ 39/45 $ \\
              & b4 & $3.2\pm0.4$ & $2.12\pm0.04$ & $30.3\pm0.6$ & $6.50\pm0.14$ & $102$ & $8.1\pm0.5$ & $16.4\pm1.3$ & $ 33/44 $ \\
\hline
4U\th 1735-44 & b1 & $2.6$ & $2.37\pm0.07$ & $15.3\pm1.3$ & $6.98\pm0.14$ & $104$ & $7.0\pm0.4 $ & $21.6\pm0.7$ & $ 44/42 $ \\
              & b2 & 2.6 & $2.20\pm0.09$ & $17.3\pm1.1 $ & $6.74\pm0.14 $ & $105$ & $7.7\pm0.4 $ & $22.4\pm0.8 $ & $ 48/41 $ \\
              & b3 & 2.6 & $2.27\pm0.07 $ & $30\pm2$ & $6.85\pm0.17 $ & $106$ & $7.4\pm0.6 $ & $25.3\pm1.2 $ & $ 27/41 $ \\
              & b4 & 2.6 & $2.31\pm0.07 $ & $38\pm2$ & $6.4\pm0.3$ & $63$ & $8.7\pm0.9 $ & $22.2\pm1.1$ & $ 30/41 $ \\
\hline
SLX\th 1735-269& i1 & $1.5$ & $0.81\pm0.07$ & $0.85\pm0.13$ & $6.8\pm0.3$ & $196$ & $77\pm20$ & $0.40\pm0.05$ & $ 27/36$ \\
               & b1 & 1.5 & $2.08\pm0.06$ & $3.9\pm0.3$ & $6.74\pm0.15$ & $79$ & $6.1\pm0.6$ & $2.90\pm0.19$ & $ 34/42 $ \\
\hline
4U\th 1820-30 & i1 & $0.3$ & $2.46\pm0.06$ & $10.9\pm0.2$ & $6.76\pm0.16 $ & $65$ & $15.3\pm0.7$ & $7.90\pm0.15$ & $ 48/48 $ \\
              & b1 & $0.3$ & $2.45\pm0.03$ & $23.8\pm1.0$ & $6.65\pm0.19$ & $44$ & $6.9\pm0.3$ & $18.0\pm0.5$ & $ 38/49 $ \\
              & b2 & $0.8\pm0.3$ & $2.31\pm0.04 $ & $27.5\pm0.8$ & $6.58\pm0.19$ & $68$ & $7.6\pm0.3 $ & $23.5\pm1.2$ & $ 26/53 $ \\
              & b3 & $1.4\pm0.3$ & $2.27\pm0.03 $ & $35.8\pm0.8$ & $6.73\pm0.18$ & $66$ & $7.4\pm0.3 $ & $27.9\pm1.4$ & $ 37/48 $ \\
\hline
4U\th 1916-053 & i6 & 0.3 & $1.34\pm0.05$ & $0.64\pm0.11$ & - & - & $76_{-30}^{+67}$ & $0.41\pm0.04$ & $ 43/42 $ \\
               & i5 & 0.3 & $1.36\pm 0.04$ & $0.74\pm0.11$ & - & - & $48\pm11$ & $0.50\pm0.04$ & $ 39/45 $ \\
               & i4 & 0.3 & $1.44\pm 0.11$ & $0.70\pm0.16$ & - & - & $22\pm4$ & $0.70\pm0.07$ & $ 44/44 $ \\
               & i3 & 0.3 & $1.52\pm 0.08$ & $0.93\pm0.17$ & - & - & $23\pm5$ & $0.82\pm0.07$ & $ 24/44 $ \\
               & i2 & 0.3 & $1.66\pm 0.11$ & $1.00\pm0.10$ & - & - & $18.6\pm1.8$ & $1.01\pm0.06$ & $ 44/44 $ \\
               & i1 & 0.3 & $2.00\pm 0.14$ & $1.39\pm0.12$ & - & - & $14.4\pm1.9$ & $1.22\pm0.07$ & $ 22/45 $ \\

\noalign{\smallskip}\hline
\end{tabular}\\
\end{center}
The column density is expressed in $10^{22}$ atom cm$^{-2}$. The blackbody temperature, cut-off energy 
and line energy are given in keV. The blackbody normalization is in $10^{36}$ erg s$^{-1}$ assuming a 
distance of 10 kpc, and the cut-off power law normalization in $10^{-1}$ keV$^{-1}$ cm$^{-2}$ at 1 keV.
The line equivalent width is in units of eV. All uncertainties are at 90\% confidence. \\
\end{table*}
%
\begin{figure}
\begin{center}                                                               
\includegraphics[width=86mm,height=86mm,angle=270]{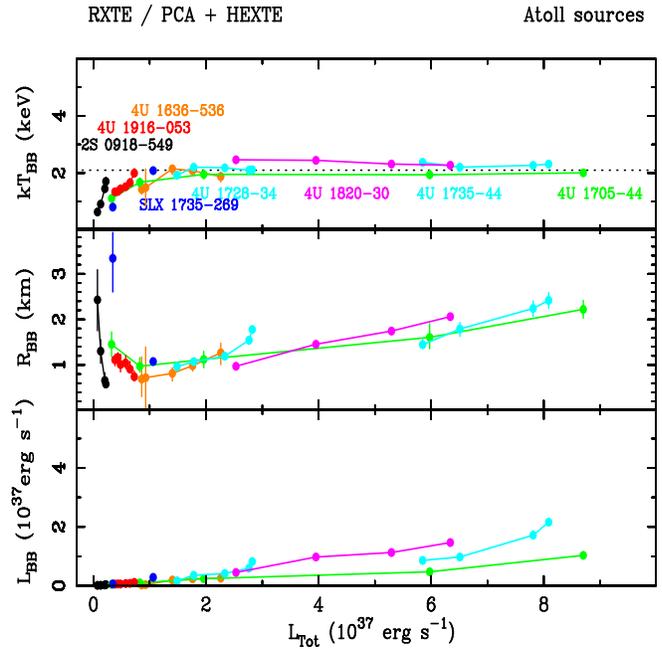}                       
\caption{Upper panel: blackbody temperature; middle panel: blackbody radius;
lower panel: bolometric blackbody luminosity.
}
\label{}
\end{center}
\end{figure}
\begin{figure}
\begin{center}                                                              
\includegraphics[width=86mm,height=86mm,angle=270]{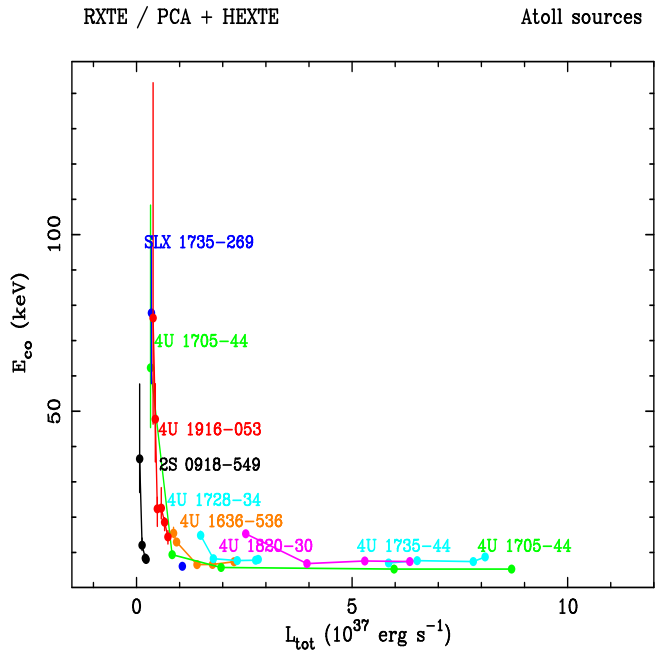}                     
\caption{High energy cut-off of the Comptonized emission as a function of total luminosity.
Error bars are plotted but in many cases are too small to be seen.
}
\label{}
\end{center}
\end{figure}
\begin{figure}
\begin{center}                                                               
\includegraphics[width=86mm,height=86mm,angle=270]{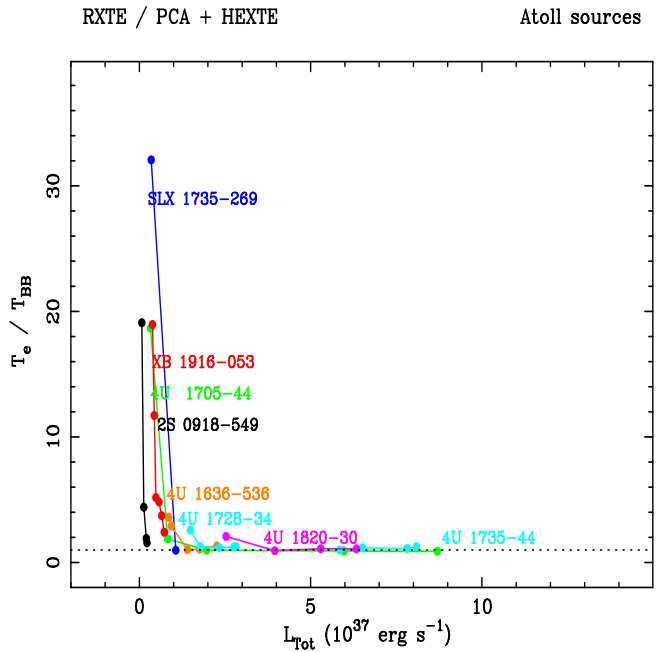}              
\caption{Ratio of Comptonizing electron temperature to neutron star blackbody temperature: 
sources shown separately. For clarity, error bars are not shown in this case.}
\label{}
\end{center}
\end{figure}

The emitting strip for a $R_{\rm BB}$ of 1 km has a half-height (i.e. above the orbital plane) $h$ of $\sim$ 100 metres.
The blackbody radius and thus $h$ increase with $L_{\rm Tot}$ consistent with findings in the {\it ASCA} survey 
(Church \& Ba\l uci\'nska-Church 2001). As will be discussed in Sect. 6, this is as expected in the model of Inogamov \& Sunyaev 
(1998, 2010) in which the accretion flow adjusts to the spin of the neutron star on the stellar surface.
In this model the accretion flow rises up the neutron star to a height that depends on the
mass accretion rates $\dot M$, i.e. $L_{\rm Tot}$.

Figure 6 (lower panel) shows the bolometric blackbody luminosity increasing with $L_{\rm Tot}$. The increase
in blackbody emission is due to the increase in blackbody radius as the temperature is constant between
10$^{37}$ - 10$^{38}$ erg s$^{-1}$, emission coming from an equatorial region of increasing height. In 4U\th 1735-44
there is a noticeable up-curving during the flare (Fig. 3).

\subsection{The Comptonized emission}

Fig. 7 shows the high energy cut-off of the Comptonized emission $E_{\rm CO}$. The diagram can be
divided into two parts. Above a critical luminosity of $\sim1 - 2\times 10^{37}$ 
erg s$^{-1}$, $E_{\rm CO}$ is low at $\approx$6 keV. This low value is continued in all of the Z-track 
sources with luminosities above $10^{38}$ erg s$^{-1}$ (Papers I -- IV) and it is a major 
feature of these sources that the spectrum does not extend to high energies. It has been
suggested that a high energy tail up to 200 keV is sometimes seen in the Z-track sources (d'Ai et al. 2007).
We have made a limited investigation of such datasets, however, with application
of the extended ADC model, a tail is not seen, i.e. the effect appear to be model-dependent, although 
it is clear that some spectral change is taking place. We stress that all our work shows that 
the high energy cut-off in the Z-track sources is low at $\sim$5 keV, the spectrum not extending to 
high energies.

At luminosities 
below the critical value, $E_{\rm CO}$ rises dramatically to tens of keV, reaching 76 keV in the case 
of XB\th 1916-053 in the observation analysed. Previous investigation of this source using {\it BeppoSAX} 
also revealed a high cut-off energy of 80$\pm$ 10 keV (Church et al. 1998b). 
The increasing errors in $E_{\rm CO}$ are, of course, due to the decreasing high energy source flux and
decreasing instrument response at energies approaching 100 keV.

The band 1 -- 30 keV was chosen for $L_{\rm Tot}$ as allowing direct comparison
with our work on the Z-track sources. Above the critical luminosity, there is very little flux beyond
30 keV. Below the critical value there may be typically 6\% of the flux above 30 keV so if data were re-plotted
using a wider band to include flux above 30 keV, the low luminosity points would be shifted only slightly
and the figure not changed substantially.

The spectra were well-fitted using a single cut-off energy, as were the {\it BeppoSAX} spectra (below).
If there was a range of electron temperatures in the ADC, a rounding of the knee would be expected, and there
is no evidence for this suggesting that the temperature of the ADC does not vary strongly radially.

The cut-off energy is determined by the maximum electron energy in the ADC plasma so that the electron temperature 
$T_{\rm e}$ can be derived from $E_{\rm CO}$. In Fig. 8, the ratio $T_{\rm e}/T_{\rm BB}$ is shown.
For a high optical depth to electron scattering $3\,kT_{\rm e}$ = $E_{\rm CO}$ while for low optical 
depth  $kT_{\rm e}$ = $E_{\rm CO}$ (Petrucci et al. 2001). 
In Fig. 8, high optical depth is assumed, i.e. the cut-off energy
is divided by 3 to give the electron temperature. It is clear that in 
sources more luminous than $\sim$1 - 2$\times 10^{37}$ \erg, the ratio is close to unity suggesting this 
assumption is correct. There is a clear indication of equality between the ADC temperature and 
the neutron star temperature, i.e. thermal equilibrium, assuming the temperature equality is not accidental
which seem unlikely. In the island state the equality breaks down and the ADC
becomes much hotter than the neutron star. For an ADC heated by the neutron star
it is impossible for this heating to produce ADC electron temperatures much higher than 1 - 2 keV,
the neutron star temperature, suggesting there is an additional unknown heating process in the ADC. 
It is not clear whether the optical depth becomes low at these low luminosities; if it does the ratio
becomes even larger. 
From Comptonization theory, for the case that the power law index remains constant (Seifina et al. 2011), 
it is expected that an increasing electron temperature means a decreasing optical depth
(Titarchuk \& Lyubarskij 1995).
The two points for the {\it Spacelab 2} source SLX\th 1735-269 shown in blue are joined by a straight line to make 
the points more obvious lacking intermediate points, although the variation is not likely to be linear.

Seifina et al. (2011, 2012) and Titarchuk et al. (2013) determine values of electron temperature, directly
from their fitted model, of 2 - 3 keV in 4U\th 1728-34, GX\th 3+1 and 4U\th 1820-30 when the sources
had higher luminosities, consistent with our results for the cut-off energy in Fig. 7 for luminosities
above the critical value.

In Fig. 9, the ratio is shown again with additional data taken from our analyses of the Z-track sources 
including data for the Cygnus\th X-2 
like sources GX\th 340+0, GX\th 5-1 and Cyg\th X-2 (Papers I -- III) and the 
Sco\th X-1 like sources \hbox{Sco\th X-1,} GX\th 349+2 and GX\th 17+2 (Paper IV).
The behaviour of the atoll sources merges with that of the Z-track sources
such that in the banana state and in the Z-sources the ADC electron temperature  
equals the neutron star temperature, consistent with the luminosity being always above the critical value.

We might expect to see a Z-shape in the Z-track sources in Fig. 9 but do not. In the Cyg-like sources
$kT_{\rm e}$ derived from $E_{\rm CO}$ varies about linearly with $kT_{\rm BB}$ on the NB
with little variation on the HB and FB. Thus the temperature ratio is about constant. In the Sco-like
sources, both temperatures are relatively constant, again giving a constant ratio.

\begin{figure}
\begin{center}                                                               
\includegraphics[width=86mm,height=86mm,angle=270]{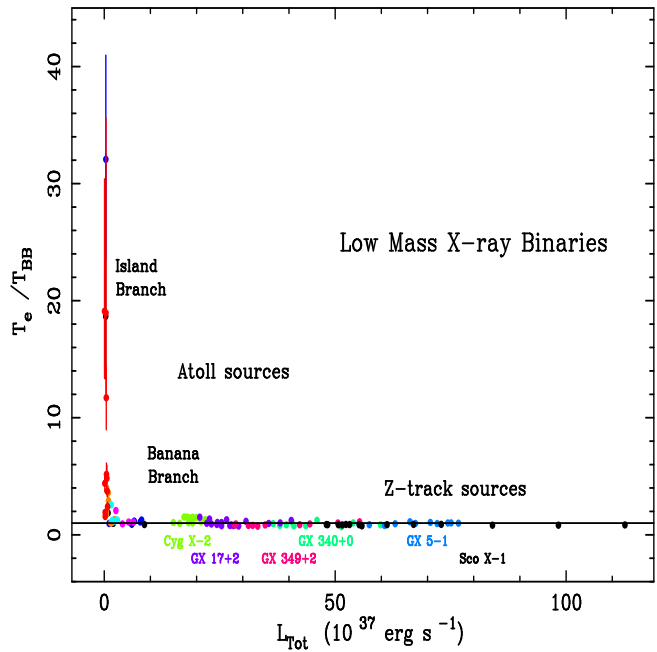}               
\caption{Atoll sources and Z-track sources: ratio of Comptonizing electron   
temperature to neutron star blackbody temperature. The solid line is drawn   
at a ratio of unity}
\label{}
\end{center}
\end{figure}


The first part of this paper based on the {\it RXTE} survey of atoll sources has demonstrated
that the high energy cut-off of the Comptonized spectrum increases sharply below 
$\approx 10^{37}$ erg s$^{-1}$, as shown in Fig. 7, demonstrating an increase of $T_{\rm e}$
in the Comptonizing ADC plasma. Thus the observed characteristics of the island state are due 
to a hot ADC while in the banana state the ADC is much cooler.

Most interesting, is the equality of $kT_{\rm e}$ and the blackbody
temperature $kT_{\rm BB}$ in the banana state, i.e. at luminosities above $\approx 10^{37}$ erg s$^{-1}$
i.e. thermal equilibrium between the neutron star and ADC. Taking $kT_{\rm e}$
to be 1/3 of $E_{\rm CO}$ produces the result that $T_{\rm e}$ = $T_{\rm BB}$
indicating that the assumption of high optical depth is justified.

The smooth transition between the banana state and the Z-track sources shown in Fig. 9 exhibiting the 
equality $T_{\rm e}$ = $T_{\rm BB}$ will suggest how the states of the atoll sources correspond, or not,
with the branches of the Z-track sources, as discussed later.

In Sect. 4, we present results of the second survey carried out using high quality
broadband {\it BeppoSAX} data which also reveals a clear systematic dependence of ADC electron temperature
on  the neutron star temperature.

\section{The BeppoSAX survey}

Table 3 shows the observations used in this investigation. The majority of these are {\it BeppoSAX}
observations analysed specifically for the present paper; in addition two other {\it SAX} observations
of the sources XB\th 1323-619 and XB\th 1916-053 previously published by us are included. These sources
with luminosities of only a few $10^{36}$ erg s$^{-1}$ extend the survey to cover more than two decades
in luminosity.

\tabcolsep 5 mm
\begin{table}                                                                   
\caption{The {\it BeppoSAX} observations in order of increasing X-ray luminosity
in the band 1- 30 keV.}
\begin{center}
\begin{tabular}{llrr}
\hline \hline \\
$\;\;$source& date & durn. & ref. \\
                  && (ksec)\\
\noalign{\smallskip\hrule\smallskip}

XB\th 1323-619    &1997, Aug 22  &119  & 2\\
XB\th 1916-053     &1997, Apr 27  &80   & 3\\
4U\th 1702-429    &2000, Sep 23  &81   & 1\\
4U\th 1543-624    &1997, Feb 21  &29   & 1\\
4U\th 1636-536    &1998, Feb 24  &71   & 1\\
4U\th 1705-44     &2000, Aug 20  &97   & 1\\
GX\th 349+2       &2001, Feb 12  &14   & 1\\
GX\th 340+0       &1998, Feb 16  &42   & 1\\
\noalign{\smallskip}\hline
\end{tabular}\\
\end{center}
References: 1: the present work; 2: Ba\l uci\'nska-Church et al. 1999; 3: Church et al. 1998b.
\end{table}

We used all four {\it SAX} narrow field instruments giving a very wide band
and spectral fitting was carried out to all four instruments simultaneously. Individual spectra 
were obtained from the Low Energy
Concentrator Spectrometer (LECS: 0.1 - 10 keV; Parmar et al. 1997), the Medium Energy Concentrator Spectrometer
(MECS; 1.3 - 10 keV; Boella et al. 1997), the High Pressure Gas Scintillation Proportional Counter
(HPGSPC; 5 - 120 keV; Manzo et al. 1997) and the Phoswich Detection System (PDS; 15 - 300 eV;
Frontera et al. 1997). The MECS consists of three grazing incidence telescopes with imaging gas scintillation
proportional counters in the focal plane. The LECS is the same, except for an ultra-thin window and a drift-less
configuration to extend the low energy response to 0.1 keV. The LECS and MECS have fields of view of 37$\arcmin$
and 56$\arcmin$, respectively.
The non-imaging HPGSPC consists of a single unit with a collimator rocked on and off source to allow background
\tabcolsep 2 mm
\begin{table*}  
\begin{center}
\caption{Results of {\it BeppoSAX} analysis: neutron star blackbody temperatures, Comptonization cut-off energies and temperature ratios.
Column densities are in units of $10^{22}$ atom cm$^{-2}$. In the case of the higher luminosity sources,
fitting was carried out typically in the energy ranges LECS: 0.1 - 5.0 keV; MECS: 1.65 - 10 keV; HPGSPC: 
7.0 - 34 keV; PDS 15.0 - 40 keV. Higher energies could not be used as the spectra became dominated by background.
However, for lower luminosity sources, energies up to 200 keV could be used, see text.
Parameter uncertainties are given at 90\% confidence.}
\begin{tabular}{llrllrll}
\hline \hline \\
$\;\;$ source &$N_{\rm H}$& $kT_{\rm BB}$&$E_{\rm CO}$&$\Gamma$&$\chi^2$/d.o.f.&$E_{\rm CO}/kT_{\rm BB}$& $E_{\rm CO}/3kT_{\rm BB}$\\
                        &&   (keV)                  &  (keV)                   &&&                            &                    \\
\noalign{\smallskip\hrule\smallskip}
XB\th 1323-619  &3.88$\pm$0.16 & 1.77$\pm$0.25   & 44$\pm$5        & 1.48$\pm$0.01 & 401/387  &  24.9$\pm$4.5      &  8.3$\pm$1.5  \\
4U\th 1702-429  &1.8$\pm$0.1   & 1.02$\pm$0.22   & 57$\pm$9        & 1.77$\pm$0.08 & 504/508  &  56$\pm$15         & 18.6$\pm$5.0  \\
XB\th 1916-053   &0.32$\pm$0.02 & 1.62$\pm$0.05   & 80$\pm$10       & 1.61$\pm$0.01 & 537/520  &  49.4$\pm$6.4      & 16.5$\pm$2.1  \\
4U\th 1543-624  &0.33$\pm$0.04 & 1.80$\pm$0.04   & 6.3$\pm$0.2     & 1.87$\pm$0.10 & 476/452  &   3.50$\pm$0.14    & 1.17$\pm$0.05 \\
4U\th 1636-536  &0.60$\pm$0.05 & 2.27$\pm$0.04   & 5.9$\pm$0.4     & 1.65$\pm$0.10 & 479/462  &   2.60$\pm$0.18    & 0.87$\pm$0.06 \\
GX\th 349+2     &1.6$\pm$0.4   & 1.63$\pm$0.04   & 5.2$\pm$1.4     & 1.49$\pm$0.54 & 208/197  &   3.19$\pm$0.86    & 1.06$\pm$0.29 \\
4U\th 1705-44   &2.3$\pm$0.2   & 1.78$\pm$0.15   & 6.0$\pm$0.3     & 1.39$\pm$0.23 & 409/393  &   3.37$\pm$0.32    & 1.12$\pm$0.11 \\
GX\th 340+0     &7.4$\pm$0.3   & 1.61$\pm$0.12   & 5.1$\pm$0.6     & 1.36$\pm$0.30 & 404/435  &   3.17$\pm$0.44   & 1.06$\pm$0.15 \\
\noalign{\smallskip}\hline
\end{tabular}\\
\end{center}
\end{table*}
\begin{figure*}                                                                  
\begin{center}
\includegraphics[width=74mm,height=74mm,angle=270]{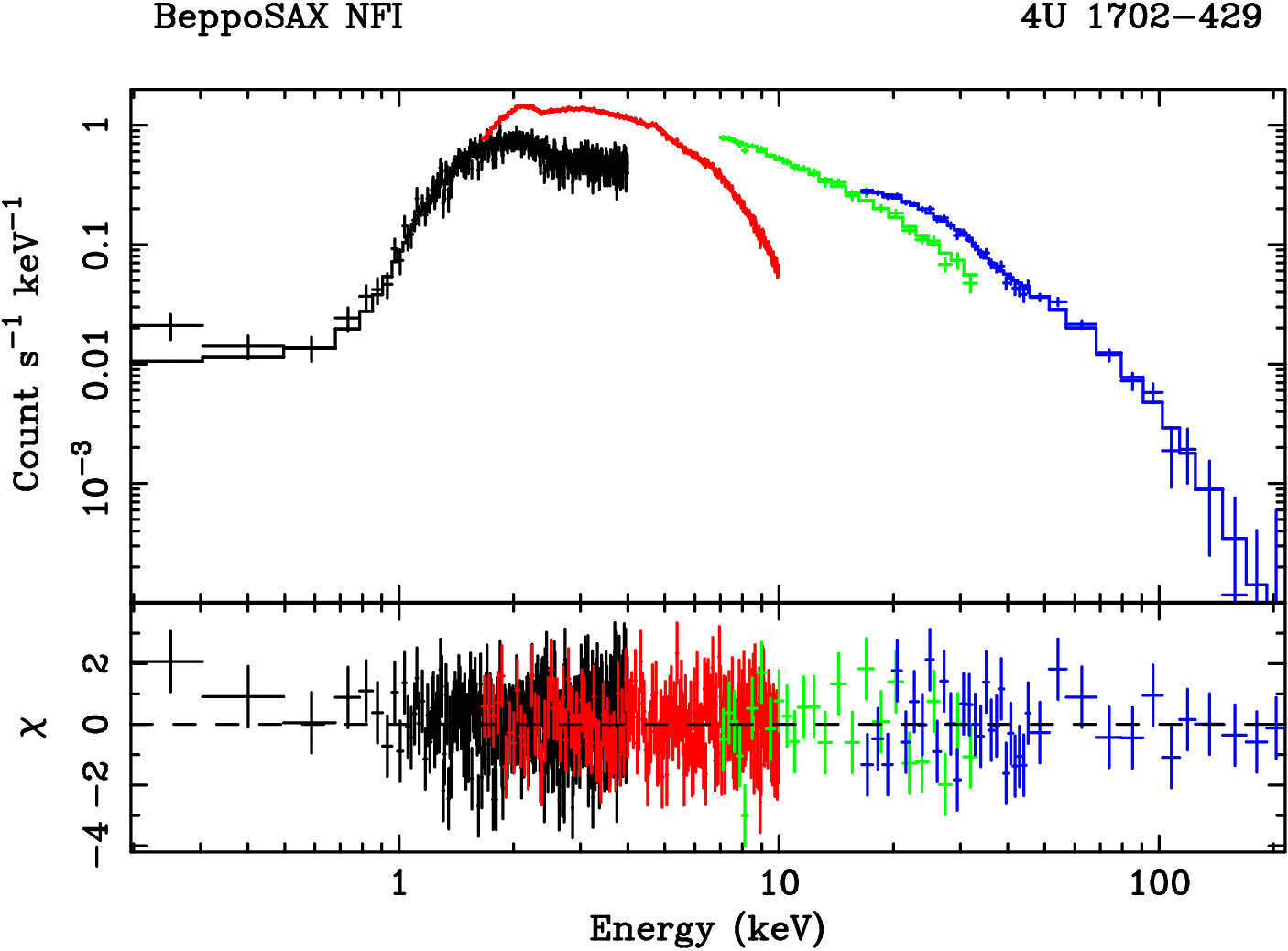}
\includegraphics[width=74mm,height=74mm,angle=270]{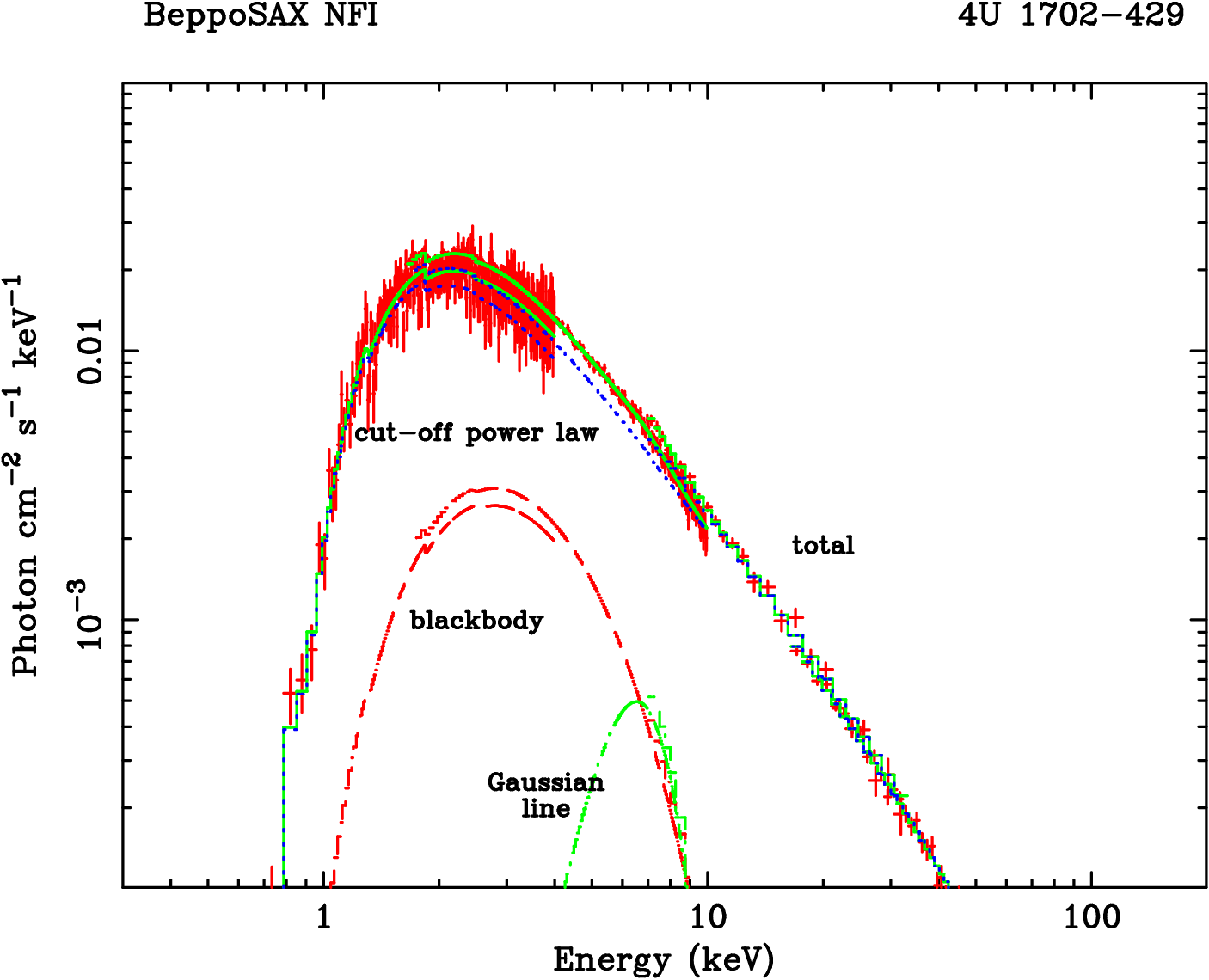}
\caption{Left: unfolded data and residuals for best-fit to {\it BeppoSAX} observation of 4U\th 1702-429:
right: folded data with individual continuum components plus a Gaussian Fe line. {\bf In this source
the spectrum extends to high energies.}}
\end{center}
\end{figure*}
determination. Similarly the PDS is non-imaging and consists of four units arranged in pairs each with a collimator
so that rocking of the collimators allows determination of the background. 

Data were selected for elevation above the Earth's limb of more than 4$\degmark$. The
{\it SAXDAS} 1.3.0 analysis package was used. LECS and MECS data were extracted from a circular region in  the
image centred on the source using radii of 8$\arcmin$ and 4$\arcmin$, respectively.
Background data for lightcurves and spectral analysis were obtained from a blank sky image.

Lightcurves were obtained from the MECS and the other instruments, and in many of the observations, there was 
little variability during the exposure and a single spectrum was extracted in a band of intensities 
to exclude any variability. In the Z-track source GX\th 349+2, some Z-track movement occurred 
and data were selected during a period of constant count rate to give a single spectrum. The sources
XB\th 1323-619 and XB\th 1916-053 display both X-ray bursting and dipping, all traces of which
were removed by selecting on time. Simultaneous fitting of the four instruments was done applying a multiplicative
factor to each instrument to allow for uncertainty in inter-instrument calibration. This
was set to unity for the MECS and determined as part of spectral fitting for the other instruments.
It was checked that spectral fitting gave acceptable values for these.

LECS data were re-binned to a minimum of
30 counts per bin, and MECS data to 40 counts per bin to allow use of the $\chi^2$ statistic. LECS and MECS
data were only used in the ranges 0.1 - 5.0 keV and 1.65 - 10 keV, respectively, where the instrument responses
are well established. The HPGSPC and PDS data were binned appropriately and used typically in the bands
7 - 34 keV and 15 - 40 keV; higher energies were excluded where the flux had fallen to the level
of the background. In 4U\th 1702-429, XB\th 1916-053 and XB\th 1323-619, the spectra were much harder
and the PDS could be used up to 200 keV. It was not necessary to add systematic errors. Because the observations were short 
without marked variation 
in intensity, hardness-intensity diagrams show each source in a localised region so that it is not 
indicated which branch or state the source was in.

The Extended ADC model was applied in the form discussed previously. In the case of {\it BeppoSAX} data
it is possible to determine the power law index of the Comptonized emission
which is usually not possible with {\it RXTE}. 
%
\begin{figure*}                                                                  
\begin{center}                                                                   
\includegraphics[width=110mm,height=160mm,angle=270]{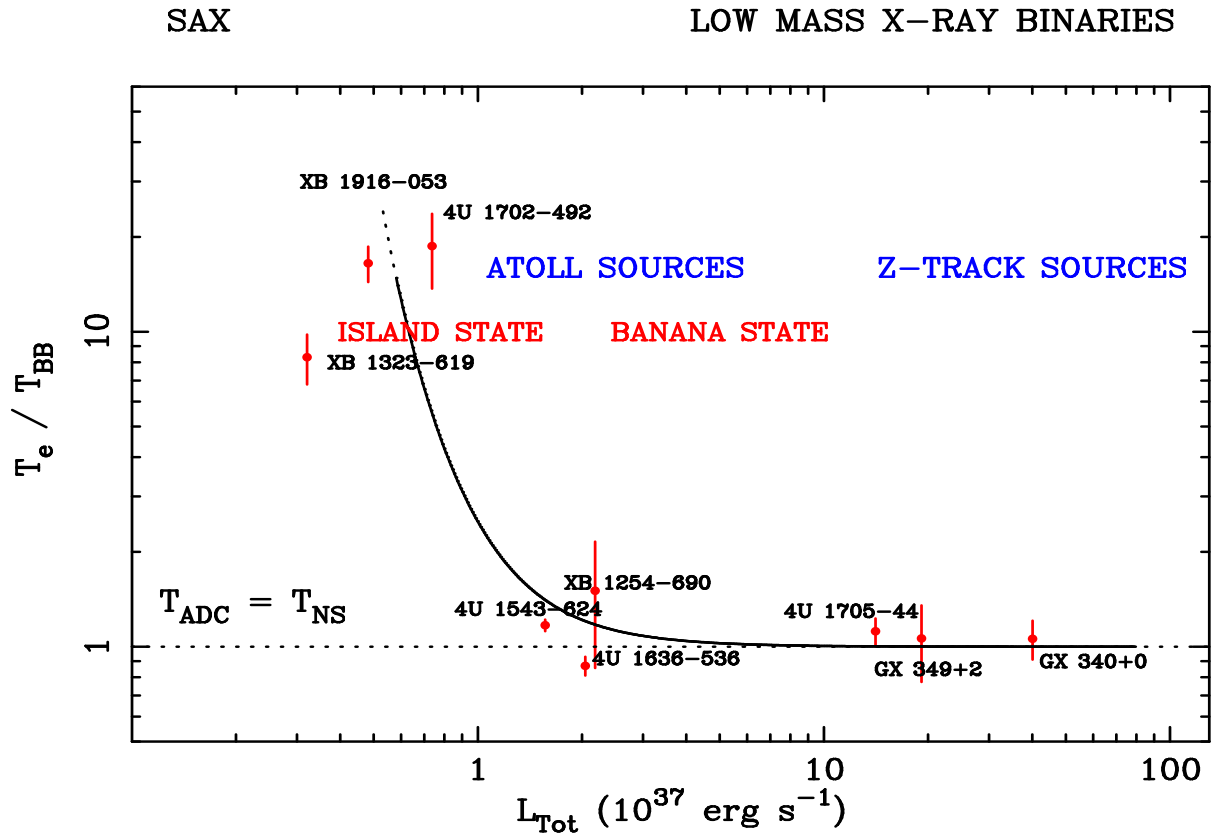}
\caption{The ratio of ADC electron temperature to neutron star blackbody temperature from
broadband spectral fitting: showing the nature of the island and banana states in atoll sources
and how these merge with the Z-track sources. A point is added from our previous {\it RXTE} observation
of XB\th 1254-690 to help define the curve drawn through the data.}
\end{center}
\end{figure*}
In cases where fitting the continuum model revealed excesses in the residuals at 
$\sim $6 keV, a Gaussian line was added to the model (in the sources 4U\th 1543-42, GX\th 349+2, 
4U\th 1702-492 and 4U\th 1705-44). This modelling provided good fits to the broadband spectra; the best-fit 
in the case of 4U\th 1702-429 is shown in Fig. 10 as a folded spectrum with residuals (left) and as an 
unfolded spectrum (right). The quality of fit was good with $\chi^2$/d.o.f. = 504/508; a broad iron line was 
detected at 6.5 keV with equivalent width 300 eV. In the sources where an Fe line was detected,
the line energy was typically $\sim$6.4 keV and
the equivalent width varied between 40 eV (4U\th 1543-42) and 300 eV (4U\th 1702-429).
%

The spectra were well fitted using a cut-off
power law for the Comptonized emission, i.e. with a single cut-off energy $E_{\rm CO}$. If there was an
appreciable range of electron temperatures in the plasma a marked rounding of the knee would result;
a factor of 10 in $T_{\rm e}$ would spread the cut-off over a similar range of energies. There is thus 
evidence that the temperature of the ADC does not vary strongly radially.

\section{Survey 2: Results}

Spectral fitting results are shown in Table 4 including the blackbody temperature $kT_{\rm BB}$ and the cut-off
energy $E_{\rm CO}$.
Also shown are the ratios $E_{\rm CO}/kT_{\rm BB}$ and $E_{\rm CO}/3\, kT_{\rm BB}$ 
corresponding to the ratio $kT_{\rm e}/kT_{\rm BB}$ in the two cases: i) low optical depth and ii) 
high optical depth (Petrucci et al. 2001). 

At higher luminosities, the first of these ratios is close to 3, and the 
second ratio close to 1 suggesting that the medium is optically thick to electron scattering
and that the ADC electron temperature equals the neutron star temperature. 

At lower luminosities, the ratios depart from unity due to the much higher cut-off 
energy. The temperature ratio
is at least 10 times higher than at higher luminosities whether we assume high or low optical depth,
and it is not clear which assumption is correct.

In Fig. 11 we show the ratio derived assuming high optical 
depth for all points to be consistent with the {\it RXTE} survey (Figs. 8, 9).
Given the small number of sources 
with luminosities of a few $10^{36}$ erg s$^{-1}$, we cannot be certain whether the temperature ratio continues 
to increase, or goes through a maximum as suggested by the point for XB\th 1323-619. To help define the
source behaviour better at intermediate luminosities, a point was added for XB\th 1254-690
from our {\it RXTE} observations of this source (Smale et al. 2002).

Clearly sources with high electron temperature are in the island state. Similarly as the temperature 
ratio falls towards unity at higher luminosities, and the spectral softness indicates the banana state. The transition 
between the states takes place at $\approx 1 - 2\times 10^{37}$ erg s$^{-1}$. Two of the sources shown are Z-track 
sources: GX\th 349+2 and GX\th 340+0 allowing us to extend Fig. 11 beyond the range of the atoll sources. The banana 
state data merge well with the Z-track data both having a temperature ratio of unity. 

\section{Discussion}

\subsection{A Model for the Atoll sources}

The results presented are based on the use of a particular model: the extended ADC model. However, the ability 
to explain the dipping sources, the  Z-track sources and now the Atoll sources in a convincing way is strong support
for the model.

The results of both surveys suggest a rather straightforward explanation.
In all LMXB more luminous than $\sim 1 - 2\times 10^{37}$ erg s$^{-1}$, i.e. atoll sources in 
the banana state, GX atolls and Z-track sources, there is thermal equilibrium between the neutron star 
and the extended ADC. This will be radiative and may be due to efficient illumination 
of the ADC by the central source. This would also imply that strong variations of temperature across the ADC 
may not be expected and the lack of spreading in energy of the knee in the high energy cut-off 
is consistent with this. As previously thought, position on the banana track is determined by $\dot M$, 
i.e. the total luminosity.

Atoll sources with luminosities less than $\approx 1 - 2\times 10^{37}$ erg s$^{-1}$ are identified with 
the island state and the hardness of this state is explained by the increase in cut-off energy. In the island
state, thermal equilibrium breaks down and the ADC temperature rises above the blackbody temperature of the 
neutron star to tens of keV. 

In our picture, the accretion disk and extended ADC are illuminated and heated by the neutron star 
and cooled by Compton cooling. There will be energy balance between these. At low mass accretion rates
the optical depth will become small in the Comptonizing ADC, so that Compton cooling becomes inefficient 
and the temperature rises. However, the temperature cannot rise higher than the temperature of the heating 
source: the neutron star. Thus the temperature of the Comptonizing region cannot exceed $\sim$2 keV by 
this heating alone.

The Eastern model with a Comptonizing region at the inner accretion disk, is inconsistent with the evidence
for an extended Comptonizing ADC. However, that approach would also require an additional
source of heating.  Historically there were attempts to explain the hard, low state of 
\hbox{Cygnus\th X-1}, for example, via a hot inner region of the accretion disk. Standard disk theory cannot produce 100 keV 
photons, since, for example, for a mass accretion rate in a LMXB corresponding to 10$^{37}$ \erg, the peak thin disk 
temperature is less than 2 keV. Shapiro et al. (1976) proposed a very hot inner disk produced by instabilities. 
Titarchuk, Lapidus \& Muslimov (1998) proposed quasi-spherical accretion with all gravitational energy deposited 
in an inner centrifugal barrier region and obtained a temperature of 
14 keV when Compton cooling is ineffective. 

Thus it is clear that there must be an additional heating mechanism capable of increasing the temperature 
of the Comptonizing region at low mass accretion rates. Our results show that the neutron star 
determines the ADC temperature at luminosities greater than the critical value and an additional
heating mechanism must be effective at lower luminosities. This is reminiscent of the solar corona in which it is 
impossible for the high coronal temperatures to be caused by photons from the photosphere, proving that 
there must be an additional heating mechanism.


A proposal for the Z-track sources has been made based on observations of the transient source 
XTE\th J1701-462 which appeared firstly as a Z-track source, then finally as an atoll source 
as it faded. Lin et al. (2009) and Homan et al. (2010) claimed that the initial Cyg-like behaviour
became Sco-like as the luminosity fell and that Cyg/Sco like type depended on luminosity. However, 
there are several strong objections to this model. Firstly, the claimed strong flaring, on which 
identification with Sco-like was based, was exaggerated by choosing an energy band of 8.6 -- 18 keV 
in which flaring appears much stronger than in a band such as 1 -- 20 keV. Our analysis producing 
a {\it RXTE} ASM lightcurve in the band 2 -- 12 keV did not show strong flaring (Paper IV). The claim 
that type depends on luminosity is not supported by our work which shows 
without doubt that the luminosity range spanned by the three Cyg-like sources Cygnus\th X-2, GX\th 340+0 
and GX\th 5-1 is the same as that spanned by the three Sco-like sources (Paper IV). Previously, 
a systematic difference between the two groups has not been found.
In Lin et al. (2009) it was suggested that $\dot M$ does 
not vary around the Z-track and they suggested rather speculative mechanisms producing the three 
branches such as the FB being caused by instabilities. This latter conflicts with the 
evidence we have presented in Papers I -- IV that flaring begins close to the critical $\dot m$
when nuclear burning of accumulated accretion material on the neutron star becomes unstable.

\subsection{How Atoll source states correspond to Z-track source states: a Unified Model for LMXB}

We relate here the proposed explanation of the island and banana states to that 
proposed for the Z-track sources (Papers I - IV). In that explanation, the mass accretion 
rate $\dot M$ was low at the Soft Apex between the normal branch and the flaring branch, but 
increased on the NB giving increased X-ray luminosity. This corresponds to the banana state in 
atolls, where increasing $\dot M$ gives increasing luminosity.

On the FB of the Z-sources, $L_{\rm ADC}$ remained constant indicating that as this was a large fraction of 
$L_{\rm Tot}$, $\dot M$ was constant while the increased total luminosity was due to increase in the
blackbody emission by energy release on the neutron star. Moreover, in the Cyg-like
sources, at the soft apex, measured values of the mass accretion rate per unit area of the neutron star $\dot m$ 
agreed well with the critical theoretical value $\dot m_{\rm ST}$ below which nuclear burning on the
surface is unstable (e.g. Bildsten 1998) and it was proposed that flaring consisted of unstable burning.


The atoll sources have much smaller $\dot M$, so are remote from $\dot m_{\rm ST}$ so cannot have the 
third basic state: the flaring state of the Z-track sources. However, unstable burning takes place in the form 
of X-ray bursting. Thus in the atoll sources (excluding the luminous GX atolls), there 
are two basic states, not three. However, flaring has been observed
in the sources XB\th 1624-490 and XB\th 1254-690 (Ba\l uci\'nska-Church et al. 2001; Smale et al. 2002) which
have luminosities in the GX atoll range so that these sources are more similar to the Z-track sources.

Conversely, the island state has no equivalent in the Z-track sources.
At low luminosity there is a strong increase in high energy cut-off $E_{\rm CO}$ due to the elevated
electron temperature that makes the spectrum much harder. In the Z-track sources $E_{\rm CO}$ is always 
low at a few keV and the variation in $E_{\rm CO}$ on the NB is very limited, e.g. from 5.5 - 7 keV in 
Cyg X-2. Thus the Z-track sources do not have this branch.

Finally, the Horizontal Branch in the Z-sources is not well understood, but our previous work
(Paper I) shows a high neutron star temperature ($>$ 2 keV) following a period of increased mass accretion. 
Eventually the neutron star must cool as the source returns to the soft apex. In 
the atoll sources, return along the banana state to lower luminosities is not accompanied by cooling (Fig. 6)
so a different track is not expected.

\subsection{The systematic change of neutron star emission with luminosity}

In the island state, the temperature $kT_{\rm BB}$ increases with increasing luminosity, i.e. $\dot M$ (Fig. 6) 
as would be expected. At the lowest luminosities, the blackbody radius $R_{\rm BB}$ increases to several km, 
and the reason for this is not known. However, the decreasing temperature will soften the spectrum.
In the lowest luminosity sources, the Extreme Island State is seen in colour colour diagrams as a
decrease in soft colour intensity at constant hard colour. It thus seems very likely that this is due to
the decreasing temperature.

At the lowest luminosities of the banana state in our surveys, the blackbody radius $R_{\rm BB}$ is small at $\sim$ 1 km
equivalent to a small equatorial emitting area on the neutron star. With increasing luminosity and $\dot M$, the emitting
area increases, $R_{\rm BB}$ rising to a few km in the GX atolls at luminosities $>$ $5\times 10^{37}$ 
erg s$^{-1}$. In the Cyg-like sources in the lowest accretion state (the Soft Apex) $R_{\rm BB}$ is 10 - 12 km 
(Papers I - III). Thus with increasing luminosity as we move from the atolls to the Z-track sources, the emitting 
area increases until the whole star is emitting. This is consistent with the model of Inogamov \& Sunyaev (1999, 2010) 
in which the boundary layer between disk and star is located on the star, accretion flow rising on the spinning star
to a height depending on $\dot M$ till the whole star is emitting at luminosities approaching the
Eddington limit.

The temperature $kT_{\rm BB}$ remains constant in the banana state, which could be expected as the increasing
accretion energy is deposited on an increasing area of neutron star. In the Cyg-like Z-track sources
$R_{\rm BB}$ cannot increase further, so increasing luminosity results in heating of the neutron star.



\section{Conclusions}

We have shown that an extended ADC description is able to provide a sensible explanation of the Island 
and Banana states in Atoll sources and how these relate to the Z-track sources, essentially a unified model 
for LMXB. We find a critical luminosity of $1- 2 \times 10^{37}$ erg s$^{-1}$ above which the ADC temperature 
equals that of the neutron star, while at lower luminosities the ADC temperature rises sharply. The model
can explain why the Atoll sources  have two basic states and the Z-track sources have three.

\thanks{
We thank the anonymous referee and Prof. Bo\.zena \hbox{Czerny} for useful discussions.
This work was supported in part by the Polish KBN grants 3946/B/H0/2008/34 and 5843/B/H03/2011/40.}


\begin{thebibliography}{}


\bibitem[]{}
Augusteijn, T., van der Hooft, F., de Jong, J. A., van Kerkwijk, M. H., \& van Paradijs, J. 1998, A\&A, 332, 561

\bibitem[]{}
Ba\l uci\'nska-Church, M., Church, M. J., Oosterbroek, T., 
et al. 1999, A\&A, 349, 495

\bibitem[]{}
Ba\l uci\'nska-Church, M., Humphrey, P. J., Church, M. J., \& Parmar, A. N. 2000, A\&A, 360, 583

\bibitem[]{}
Ba\l uci\'nska-Church, Barnard, R., Church, M. J., \& Smale, A. P. 2001, A\&A, 378, 847

\bibitem[]{}
Ba\l uci\'nska-Church, M., Gibiec, A., Jackson, N. K., \& Church, M. J. 2010, A\&A, 512, A9 (Paper III)



\bibitem[]{}
Barnard, R., Ba\l uci\'nska-Church, M., \& Smale, A. P., \hbox {Church, M. J.} 2001, A\&A, 380, 494



\bibitem[]{}
Bildsten, L. 1998, in Proc NATO ASIC 515,
The Many Faces of Neutron Stars, eds. R. Buccheri,
J. van Paradijs \& M. A. Alpar, Dordrecht-Kluwer, 419

\bibitem[]{}
Bloser, P. F., Grindlay, J. E., Kaaret, P., et al. 2000, ApJ, 542, 1000

\bibitem[]{}
Boella G., Chiappetti L., Conti G., et al, 1997, A\&AS 122, 327



\bibitem[]{}
Chevalier, C., \& Ilovaisky, S. A. 1987, A\&A, 172, 167



\bibitem[]{}
Church, M. J., \& Ba\l uci\'nska-Church, M. 1995, A\&A, 300, 441

\bibitem[]{}
Church, M. J., \& Ba\l uci\'nska-Church, M. 2001, A\&A, 369, 915

\bibitem[]{}
Church, M. J., \& Ba\l uci\'nska-Church, M. 2004, MNRAS, 348, 955

\bibitem[]{}
Church, M. J., Mitsuda, K., Dotani, T., et al. 1997,
ApJ, 491, 388

\bibitem[]{}
Church, M. J., Ba\l uci\'nska-Church, M., Dotani, T., \& Asai, K. 1998a, ApJ, 504, 516

\bibitem[]{}
Church, M. J., Parmar, A. N., Ba\l uci\'nska-Church, M., et al. 1998b, A\&A, 338, 556

\bibitem[]{}
Church, M. J., Inogamov, N. A., \& Ba\l uci\'nska-Church, M. 2002, A\&A, 390, 139


\bibitem[]{}
Church, M. J., Halai, G. S., \& Ba\l uci\'nska-Church, M. 2006, A\&A, 460, 233 (Paper I)

\bibitem[]{}
Church, M. J., Gibiec, A., Ba\l uci\'nska-Church, M., \& Jackson, N. K. 2012, A\&A, 546, A35

\bibitem[]{}
Christian, D. J., \& Swank, J. H. 1997, ApJS, 109, 177

\bibitem[]{}
D'Ai A., \.Zycki P, di Salvo T., Iaria R., Lavagetto G., \& Robba N. R. 2007, ApJ 667, 411



\bibitem[]{}
di Salvo, T., Iaria R., Burderi L., \& Robba N. R., 2000, ApJ, 542, 1034






\bibitem[]{}
Falanga, M., G\"otz, D., Goldoni, P., et al. 2006, A\&A, 458, 21

\bibitem[]{}
Frontera F., Costa E., Dal Fiume D., et al. 1997, A\&AS 122, 371






\bibitem[]{}
Galloway, D. K., Muno, M. P., Hartman, J. M., Psaltis, D., \& Chakrabarty, D 2008, ApJS, 179, 360


\bibitem[]{}
Goldwurm, A., Vargas, M., Paul, J., et al. 1996, A\&A, 310, 857

\bibitem[]{}
Grindlay, J. E., Gursky, H., Schnopper, et al. 1976, ApJ, 205, L127
Brinkman, A. C. 1976, ApJ, 205, L127

\bibitem[]{}
Hasinger, G., \& van der Klis, M. 1989, A\&A, 225, 79




\bibitem[]{}
Hjellming, R. M., Stewart, R. T., White, G. L., et al. 1990, ApJ, 365, 681

\bibitem[]{}
Homan, J., van der Klis, M., Fridriksson, J. K., et al. 2010, ApJ, 719, 201


\bibitem[]{}
Inogamov, N. A., \& Sunyaev, R. A. 1999, AstL, 25, 269

\bibitem[]{}
Inogamov, N. A., \& Sunyaev, R. A. 2010, AstL, 36, 848

\bibitem[]{}
Jackson, N. K., Church, M. J., \& Ba\l uci\'nska-Church, M. 2009, A\&A, 494, 1059 (Paper II)

\bibitem[]{}
Jahoda, K., Swank, J. H., Giles, A. B., et al. 1996, SPIE, 2808, 59 

\bibitem[]{}
Jimenez-Garate, M. A., Raymond, J. C., \& Liedahl, D. A. 2002, ApJ, 581, 1297


\bibitem[]{}
Kaminker , A. D., Pavlov, G. G., Shibanov, Y. A., et al. 1989, A\&A, 220, 117



\bibitem[]{}
Lewin W. H. G., van Paradijs J., \& Taam R. E. 1995 in ``X-ray Binaries, eds.
W. H. G. Lewin, J. van Paradijs,  E. P. J. van den Heuvel, Cambridge.

\bibitem[]{}
Lin D., Remillard R. A., Homan J., 2009, ApJ, 696, 1257


\bibitem[]{}
Manzo G., Guarrusso S., Santangelo A., et al. 1997, A\&AS 122, 341

\bibitem[]{}
Mitsuda, K., Inoue, H., Nakamura, N., \& Tanaka, Y. 1989, PASJ, 41, 97

\bibitem[]{}
Muno, M. P., Remillard, R. A., \& Chakrabarty, D. 2002, ApJ, 568, L35

\bibitem[]{}
Parmar A.N., Martin D.D.E., Bavdaz M., et al. 1997, A\&AS 122, 309

\bibitem[]{}
Petrucci, P. O., Haardt, F., Maraschi, L., et al. 2001, ApJ, 556, 716









\bibitem[]{}
Schulz, N. S., Hasinger, G., \& Tr\"umper, J. 1989, A\&A, 225, 48

\bibitem[]{}
Schulz, N. S., Huenemoerder, D. P., Ji, L., Nowak, M., et al.
2009, ApJ, 692, L80

\bibitem[]{}
Seifina E., \& Titarchuk L., 2011, ApJ, 738, 128 

\bibitem[]{}
Seifina E., \& Titarchuk L., 2012, ApJ, 747, 99

\bibitem[]{}
Seifina E., Titarchuk L., \& Frontera F., 2013, ApJ, 766, 63

\bibitem[]{}
Shapiro, S. L., Lightman, A. P., \& Eardley, D. M. 1976, ApJ, 204, 187

\bibitem[]{}
Smale, A. P., Church, M. J., \& Ba\l uci\'nska-Church, M. 2001, ApJ, 550, 962

\bibitem[]{}
Smale, A. P., Church, M. J., \& Ba\l uci\'nska-Church, M. 2002, ApJ, 581, 1286

\bibitem[]{}
Titarchuk L., \& Lyubarskij Y., 1995, ApJ, 450, 876

\bibitem[]{}
Titarchuk L., Lapidus I., Muslimov A., 1998, ApJ, 499, 315


\bibitem[]{}
Titarchuk L., Seifina E., \& Frontera F., 2013, ApJ, 767, 160

\bibitem[]{}
van der Klis 1995, in ``X-ray Binaries'',ed. W. H. G. Lewin, J. van Paradijs, and E. P.
J. van den Heuvel. Cambridge and New York, Cambridge University Press, 1995, 252


\bibitem[]{}
van der Klis 2006, in ``Compact Stellar X-ray Sources'', eds. W. H. G. Lewin and 
M. van der Klis, Cambridge Astrophysics Series 39, Cambridge. 

\bibitem[]{}
van Straaten, S., van der Klis, M., \& M\'endez, M. 2003, ApJ, 596, 1155

\bibitem[]{}
White, N. E., \& Swank, J. H. 1982, ApJ 253, L66

\bibitem[]{}
White, N. E., Stella, L., \& Parmar., A. N. 1988, ApJ, 324, 363

\bibitem[]{}
White, N. E., Peacock, A., Hasinger, G., et al. 1986, MNRAS, 218, 129


\bibitem[]{}
Woosley, S. E., \& Taam, R. E. 1976, Nature, 263, 101

\bibitem[]{}
van Straaten S., van der Klis M., M\'endez M., 2003, ApJ, 596, 1155



\end{thebibliography}
\end{document}